\documentclass[twocolumn]{article} 
\usepackage[T1]{fontenc}
\usepackage[latin1]{inputenc} 
\usepackage{geometry}
\usepackage{graphics}


\begin{document}

\twocolumn[\hsize\textwidth\columnwidth\hsize\csname@twocolumnfalse\endcsname

\title{ARPES Line Shapes in FL and non-FL Quasi-Low-Dimensional Inorganic
  Metals}

\author{G.-H. Gweon\protect\( ^{a}\protect \), J. D. Denlinger\protect\(
  ^{a,1}\protect \), J. W. Allen\protect\( ^{a}\protect \), R.
  Claessen\protect\( ^{b}\protect \), C. G. Olson\protect\( ^{c}\protect
  \), H. H\"{o}chst\protect\(
  ^{d}\protect \)\\
  J. Marcus\protect\( ^{e}\protect \), C. Schlenker\protect\( ^{e}\protect
  \)
  and L. F. Schneemeyer\protect\( ^{f}\protect \)\\
  \emph{\normalsize \protect\( ^{a}\protect \)Randall Laboratory of
Physics, University of Michigan, Ann Arbor, MI 48109-1120, USA}\\
  \emph{\normalsize \protect\( ^{b}\protect \)Experimentalphysik II,
    Universit}\emph{\normalsize \"a}\emph{\normalsize t
    Augsburg, D-86135 Augsburg, Germany}\\
  \emph{\normalsize \protect\( ^{c}\protect \)Ames Laboratory, Iowa State
    University,
    Ames, Iowa 50011, USA}\\
  \emph{\normalsize \protect\( ^{d}\protect \)Synchrotron
    Radiation Center, University of Wisconsin, Stoughton, WI 53589, USA}\\
  \emph{\normalsize \protect\( ^{e}\protect \)Laboratoire d'Etudes des
    Propri}\emph{\normalsize \'e}\emph{\normalsize t}\emph{\normalsize
    \'e}\emph{s}
  \emph{\normalsize Electroniques des Solides -- }\\
  \emph{\normalsize CNRS, BP166, 38042 Grenoble Cedex, France}\\
  \emph{\normalsize \protect\( ^{f}\protect \)Bell Laboratories, Lucent
    Technologies, 700 Mountain Avenue, Murray Hill, New Jersey
    07974, USA}\normalsize }

\maketitle
\begin{abstract}
  {\normalsize Quasi-low-dimensional (quasi-low-D) inorganic materials are
    not only ideally suited for angle resolved photoemission spectroscopy
    (ARPES) but also they offer a rich ground for studying key concepts
    for the emerging paradigm of non-Fermi liquid (non-FL) physics. In
    this article, we discuss the ARPES technique applied to three
    quasi-low-D inorganic metals: a paradigm Fermi liquid (FL) material
    TiTe\( _{2} \), a well-known quasi-1D charge density wave (CDW)
    material K\( _{0.3} \)MoO\( _{3} \) and a quasi-1D non-CDW material
    Li\( _{0.9} \)Mo\( _{6} \)O\( _{17} \).  With TiTe\( _{2} \), we
    establish that a many body theoretical interpretation of the ARPES
    line shape is possible. We also address the fundamental question of
    how to accurately determine the} \textbf{\normalsize k}{\normalsize \(
    _{F} \) value from ARPES. Both K\( _{0.3} \)MoO\( _{3} \) and Li\(
    _{0.9} \)Mo\( _{6} \)O\( _{17} \) show quasi-1D electronic structures
    with non-FL line shapes. A CDW gap opening is observed for K\( _{0.3}
    \)MoO\( _{3} \), whereas no gap is observed for Li\( _{0.9} \)Mo\(
    _{6} \)O\( _{17} \). We show, however, that the standard CDW theory,
    even with strong fluctuations, is not sufficient to describe the
    non-FL line shapes of K\( _{0.3} \)MoO\( _{3} \). We argue that a
    Luttinger liquid (LL) model is relevant for both bronzes, but also
    point out difficulties encountered in comparing data with theory. We
    interpret this situation to mean that a more complete and realistic
    theory is necessary to understand
    these data.}\\


\end{abstract}
]

\section{Introduction}

Angle resolved photoemission spectroscopy (ARPES) is one of the most
direct probes of the electronic structure of solids. By directly measuring
single-particle excitation spectra as a function of momentum and energy, it
can determine the most basic quantities of condensed matter physics, e.g.\ 
the band structure, Fermi surface (FS) and electronic gap opening.
Furthermore, the (AR)PES line shape can give crucial information about
important ground state properties as discussed in other articles in this
volume. For technical reasons, ARPES is especially well suited to
quasi-2-dimensional (quasi-2D) and quasi-1D layered materials in which the
electron dispersion perpendicular to the cleavage surface is small. In
this case, it becomes simpler to interpret ARPES because one is primarily
concerned with momenta parallel to the surface which are conserved
quantities in the photoemission process and because the photohole line
shape is free from the finite photoelectron lifetime induced momentum
averaging effect
\cite{Bardyszewski:1985:pes-theory,Smith:1993:photo-electron-lifetime},
which can severely limit the momentum resolution along the surface normal
direction and give an added broadening of the line shape.

Quasi-low-D materials are interesting because interacting 1D systems are
fundamentally different from interacting 3D systems. In 3D, the Landau
Fermi liquid (FL) theory \cite{Laudau:1956:FL,Laudau:1957:FL} is a
well-accepted paradigm. In this theory, a system of strongly interacting
electrons (and holes) is viewed as that of weakly interacting
quasi-particles with enhanced masses. In 1D, a FL is completely unstable.
First, forward scatterings between electrons give rise to a Luttinger
liquid (LL) \cite{Haldane:1981:LL}, in which a quasi-particle no longer
exists and spin and charge collective excitations completely describe the
low energy physics. We will discuss the LL further in Section \ref{sec: li
  bronze}.  Second, backward scattering of an electron from one FS to the
other leads to spontaneous charge density wave (CDW) formation
\cite{Peierls:1955:quantum-theory-of-solids-book}, which opens up a gap at
\( E_{F} \), making the material become an insulator \cite{SDW}. The
standard CDW model is the Fr\"{o}lich model
\cite{Froehlich:1954:froehlich-hamiltonian}.  The mean field solution of
this model is formally equivalent to the BCS solution for
superconductivity. However, in a quasi-1D system, fluctuation effects are
expected to be very important and, e.g.\ lead to a pseudo-gap in the
normal state. We will discuss the CDW theory further in Section \ref{sec:
  blue bronze}.

Lately, interest in low-D physics seems to be expanding rapidly, partly
due to high interest in low-D artificial structures and nano-scale
materials. However, at this stage, it may be said that a proper
understanding of real low-D materials is lacking. For example, high
temperature superconductors show behaviors in photoemission, e.g.\ 
pseudo-gaps \cite{Ding:1996:Nature,Loeser:1996:Science}, signs of
critical fluctuations \cite{Feng:2000:Science} and strange normal state
line shapes \cite{Olson:1990:PRB}, which are clearly not understood
within the standard BCS theory and which still await a coherent
explanation. Similarly, CDW materials, such as the blue bronze
\cite{Gweon:1996:JPhys-CondMatt} and TTF-TCNQ \cite{Zwick:1998:PRL} show
anomalies that are not reconcilable within the standard mean-field
Fr\"{o}lich model. In studying quasi-1D materials, it seems a necessity to
learn the importance of the two phenomena inherent to 1D -- the CDW and
the LL.  Note also that a real system is never strictly 1D but always has
residual 3D couplings between chains. Therefore, a proper understanding of
3D couplings is also important. In fact, one may say that 3D couplings are
essential to understand quasi-1D materials, because (1) a finite \( T \)
CDW transition is possible only because of them, and (2) interacting
electrons strictly in 1D form a Wigner lattice
\cite{Schulz:1993:wigner-lattice} instead of the LL, due to unscreened
long range Coulomb interaction.

In this article, we show how ARPES data let us confront these difficult
but fascinating issues. Especially using a state of the art high
resolution spectrometer such confrontations become more revealing than
previously. We discuss three systems, a FL reference TiTe\( _{2} \), a
quasi-1D non-CDW, the "Li purple bronze" Li\( _{0.9} \)Mo\( _{6} \)O\(
_{17} \), and a quasi-1D CDW, the "blue bronze" K\( _{0.3} \)MoO\( _{3}
\). These three materials represent three quite different categories but,
like the high temperature superconductor (HTSC) Bi2212, all are inorganic
layered 3D crystals for which large samples are available.  Such materials
are high on the priority list for ARPES studies.  Cleaving yields high
quality surfaces of large area which are more stable for ARPES than those
of many organic low-D conductors.  Thus it is easier to obtain
reproducible bulk-representative ARPES data.  Also conventional thermal
and transport data are readily available to be correlated with the ARPES
data. Meeting all these prerequisites simultaneously is often difficult
for other kinds of low-D materials.

This paper is organized as follows. Section \ref{sec: Theory} describes
the theoretical background for ARPES and Section \ref{sec:
  Experiment} summarizes experimental conditions.
Section \ref{sec: TiTe2} describes the
FL interpretation of ARPES data for the Ti \( 3d \) band of TiTe\( _{2}
\).  The special property of the Ti \( 3d \) band that its Fermi velocity
(\( v_{F} \)) is small leads to a quite unusual situation in which the
ARPES dispersion moves \textit{across} the chemical potential, \( \mu \).
We report such data and compare the \( \mathbf{k}_{F} \) value determined
from it with values estimated by other methods used by ARPES
practitioners.  Section \ref{sec: li bronze} describes the photoemission
data of Li\( _{0.9} \)MoO\( _{17} \) and Section \ref{sec: blue bronze}
describes the photoemission data of K\( _{0.3} \)MoO\( _{3} \).  We report
the absence in Li\( _{0.9} \)MoO\( _{17} \) and the presence in K\( _{0.3}
\)MoO\( _{3} \) of a gap opening associated with their phase transitions.
We discuss non-FL line shapes found for these materials in view of
currently available LL and CDW theories.

\section{Theoretical Framework \label{sec: Theory}}

Within the sudden approximation \cite{HedinAndLundqvist:1969}, the ARPES
line shape is described, up to a matrix element factor, as
\begin{equation}
\label{IkwT}
I(\mathbf{k},\omega ,T)=\sum _{\omega ^{\prime }}f(\omega ^{\prime },T)\,
\sum _{\mathbf{k}^{\prime }}A(\mathbf{k}^{\prime },\omega ^{\prime },T)
\end{equation}
where \( f \) is the Fermi-Dirac distribution function, \( A \) is the
single particle spectral function and \( T \) is temperature. Sums over
\textbf{\( \mathbf{k}^{\prime } \)} and \( \omega ^{\prime } \) account
for the momentum and energy resolutions of the instrument, respectively,
with resolution functions implied in the summation notation for
simplicity. The energy resolution function can be obtained from the Fermi
edge of a reference sample (polycrystalline Ag, Au or Pt) and, in our
case, is found to be a gaussian function to a good approximation. The
momentum resolution function can be modeled based on geometrical
considerations. For our cases, where the band dispersion is dominant along
one direction, it can be modeled as a gaussian sum over momentum along
that direction.

In order to understand the ARPES line shape, it is quite useful to
mentally process Eq.\ (\ref{IkwT}) from right to left. The final step --
convolution in \( \omega ^{\prime } \) -- is not important for a
qualitative understanding, because its effect is ``just'' energy
broadening. The most important part in Eq.\ (\ref{IkwT}) is the spectral
function \( A \), which is by definition \( \mathrm{Im}\, G/\pi \), where
\emph{\( G \)} is the single particle Green's function. Often, \( G \) is
written as \( 1/(\omega -\epsilon (\mathbf{k})-\Sigma (\mathbf{k},\omega
)) \) where \( \epsilon (\mathbf{k}) \) is the one-electron band energy
for momentum \( \mathbf{k} \) and \( \Sigma (\mathbf{k},\omega ) \) is the
so-called self energy, which contains all the information about the
single-particle interaction physics of the system.

There are other general effects that we do not consider in this article.
First, an additional sum over momentum along the surface normal direction,
\( \mathbf{k}_{\perp} \), must be included to account for the finite
photo-electron lifetime \cite{Bardyszewski:1985:pes-theory}.  This effect
is minimized for quasi-low-D materials, as already noted in the previous
section, because then \( A \) is not
dependent on \( \mathbf{k}_{\perp} \) to a first approximation. An
estimate of an upper-bound for the line broadening due to this effect can
be made for each of our materials, as is done for TiTe$_2$
\cite{Claessen:1996:PRB}, and confirms that this effect can be safely
neglected. Another effect is the inelastic background
\cite{Shirley:1972:shirley-background}.  In general, this is quite
difficult to quantify for ARPES, and remains an important issue especially
for the HTSC's \cite{Norman:1999:PRB}. In the next section, we will see
that the TiTe\( _{2} \) data show an extremely low background. We take
this to imply the likelihood that the inelastic background is very small
within \( \sim 1 \) eV from the
chemical potential \( \mu \) for ARPES data taken with photon energy \(
\sim 20 \) 
eV on good cleaved surfaces of other samples as well.

\section{Experimental Setup \label{sec: Experiment}}

ARPES data reported in this article were obtained at the Wisconsin
Synchrotron Radiation Center (SRC). ARPES data were taken at the
Ames/Montana beamline with a 50 mm radius VSW analyzer having a \( \pm
1^{o} \) angle acceptance cone.  ARPES data with angle resolution \( \pm
0.18^{o} \) along the main band dispersion axis were obtained at the 4m
NIM line or the PGM line with a Scienta SES 200 analyzer. Angle integrated
PES data were obtained with a VG ESCALAB Mk II spectrometer (\( \pm 12^{o}
\) angle acceptance cone) in the home lab or with the Scienta SES 200
analyzer in the angle integrated mode (angle resolution \( \pm 6^{o} \)
along the main band dispersion axis). The angle resolution of the Scienta
analyzer perpendicular to the main band dispersion, $\pm 0.1^{o}$ to
$\pm 0.25^{o}$, is 
irrelevant for the discussion here. Hereafter, we will implicitly use the
relevant angle resolution as a unique identifier for the spectrometer with
which the data were taken. All samples were oriented with Laue photographs
and were cleaved \emph{in situ} with a top-post method.

\section{TiTe\protect\( _{2}\protect \) -- FL Line Shape\label{sec: TiTe2}}

TiTe\( _{2} \) is a layered compound which is a semi-metal due to the
small energetic overlap of a set of nominally Te \( 5p \) bands and one
orbitally non-degenerate Ti \( 3d \) band \cite{Claessen:1996:PRB}.
Its transport properties give no indication of any behavior lying outside
of the FL framework \cite{Koike:1983:tite2-resistivity}, and it is known
to be metallic down to the lowest measured temperature 1.1 K
\cite{Allen:1994:allen-chetty}. This physical property makes TiTe\( _{2}
\) an attractive candidate for ARPES study as a reference FL system, to
which ARPES data for exotic materials can be compared.

TiTe\( _{2} \) is a gift of nature for an ARPES line shape study. The
overall band structure of this material is now well understood
\cite{Claessen:1996:PRB} both theoretically and experimentally. There
is good agreement between the band calculation and experiment regarding
the number of bands and the character and the shape of the FS pieces,
which have also been measured by the intensity map method
\cite{Straub:1997:PRB}. What makes this material so special for ARPES
is the fact that the Ti \( 3d \) band is well isolated from other bands.
In addition, the spectra are exceptionally clean, almost entirely free of
an inelastic background signal.

\begin{figure}
  {\centering \resizebox*{0.95\columnwidth}{!}{\includegraphics{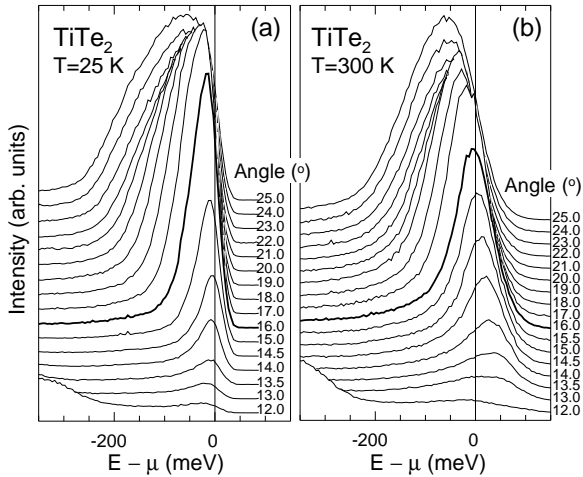}}
    \par}

\caption{ARPES data of the Ti \protect\( 3d\protect \) band of
  TiTe\protect\( _{2}\protect \) taken at \protect\( h\nu =21.2\protect \)
  eV\@. Energy and angle resolutions are 35 meV and \protect\( \pm
  1^{o}\protect \), respectively. Thick lines are for \protect\(
  \mathbf{k}=\mathbf{k}_{F}\protect \). \label{fig: TiTe2 data}}
\end{figure} 

Fig.\ \ref{fig: TiTe2 data} shows our ARPES data taken at 25K and 300K\@.
Previously, we have reported FL line shape fits of a 25 K data set
\cite{Claessen:1992:PRL,Allen:1995:JPhysChemSol}.  The new 25 K data in
Fig.\ \ref{fig: TiTe2 data} are practically identical with our previously
reported data \cite{Claessen:1992:PRL,Allen:1995:JPhysChemSol}.  Note
that the constant intensity at high binding energy, at least part of which
is due to inelastic background, is negligibly low compared to the peak
height in the data.

In previous reports \cite{Claessen:1992:PRL,Allen:1995:JPhysChemSol}, we
have shown that the line shapes at low \( T \) are described well by
Eq.\ \ref{IkwT} with a FL theory. The FL theory that we used in
Ref.\ {[}\ref{Allen:1995:JPhysChemSol}{]} is a simple phenomenological causal
theory that has the correct FL behavior at low energies and satisfies the
spectral sum rule on the global scale. As explained in
Ref.\ {[}\ref{Allen:1995:JPhysChemSol}{]} in detail, this theory involves two
poles in the Green's function, a quasi-particle pole and a ``background''
pole.  The overall line shape evolution as a function of \( |\epsilon
(\mathbf{k})| \) shows crossover from a heavy quasi-particle band
dispersion to an un-renormalized band dispersion. In the crossover region,
the two poles interfere to produce an interesting two peak line shape,
which we identified with the exceptionally broad line shape at large
angles (e.g.\ 25\( ^{o} \)). This crossover behavior is quite analogous to
the similar behaviors found in the strong electron-phonon coupling systems
\cite{Hengsberger:1999:PRL,Valla:1999:PRL,Lashell:2000:PRB}
or in the HTSC's \cite{Balatsky:1999:Science}. The main finding of the
previous fit efforts was that the quadratically falling tail at the high
binding energy side of the peak distinguishes the FL model from other
models. Here, we will focus on the temperature dependence of the line
shape.

Eq.\ \ref{IkwT} shows that \( T \) dependent line shape changes can occur
due to both \( A \) and \( f \). For TiTe\( _{2} \), which undergoes no
phase transition or crossover, the change in \( A \) is expected to be
simply a gradual increase in the line width as \( T \) increases. Without
knowing the exact \( T \) dependence of \( A \), we will first ignore it
as an approximation, and then investigate to what extent this
approximation departs from observation. The \( T \) dependence of the
Fermi-Dirac function \( f \) is simple: the step at \( \mu \) becomes
wider and flatter. The \( T \)-linear increase of the step width means
that a larger portion of \( A \) above \( \mu \) becomes visible in
photoemission.

Within this approximation, the most outstanding feature of the ARPES data
at 300 K, i.e.\ that the dispersing peak is observed across \( \mu \), can
be understood by simple considerations. Near \( \mu \), the intrinsic
quasi-particle spectral function is approximately a delta function.
However, the finite angular resolution requires that the spectral function
must be summed over a momentum window \( \Delta \mathbf{k} \) to give an
effective energy width \( \hbar v^{\prime }_{F}\Delta \mathbf{k} \) where
\( \hbar v^{\prime }_{F} \) is the peak velocity near \( \mu \). Taking
the FWHM of the \( \omega \)-derivative of \emph{\( f \)}, the width of
the step in \emph{f} is approximately \( 4k_{B}T \). An interesting case
occurs if this width is larger than the width of the peak, \( \hbar
v^{\prime }_{F}\Delta \mathbf{k} \).  In this case, a peak slightly, say
\( k_{B}T \), above \( \mu \), has most of its intensity above \( \mu \),
so that even after the multiplication by \( f \) in Eq.\ \ref{IkwT}, the
line shape shows a peak \textit{above} \( \mu \).  For the current case,
\( \hbar v^{\prime }_{F} \) is \( \sim 0.5 \) eV\AA\ and \( \Delta
\mathbf{k} \) is \( 0.07 \) \AA\( ^{-1} \), which gives an estimate of \(
\hbar v^{\prime }_{F}\Delta \mathbf{k}=35 \) meV\@. At 300 K, \(
4k_{B}T=100\mathrm{meV} \), significantly larger than 35 meV, and indeed,
we see the peak crossing \( \mu \).

\begin{figure}
  {\centering \resizebox*{0.95\columnwidth}{!}{\includegraphics{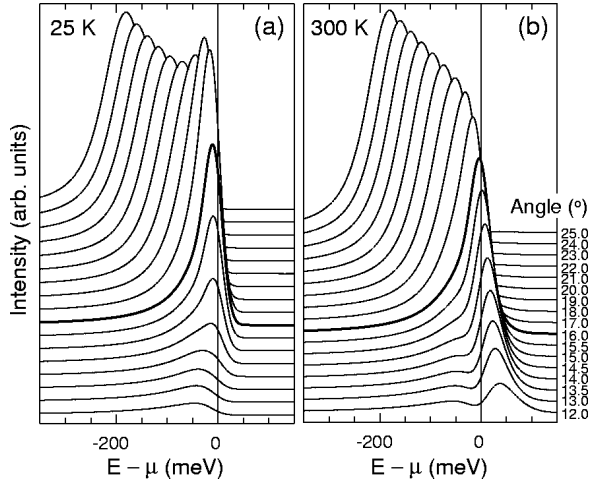}}
    \par}

\caption{FL line shape simulations for the data of Fig.\ \ref{fig: TiTe2
    data}. The model parameters \cite{Allen:1995:JPhysChemSol} are \protect\(
  Z/Q=0.4\protect \), \protect\( Q\hbar v_{F}=0.6\protect \) eV\AA, and
  \protect\( 1/\beta ^{\prime }=40\protect \) meV\@. Thick lines are for
  \protect\( \mathbf{k}=\mathbf{k}_{F}\protect \).
\label{fig: TiTe2 simulation}}
\end{figure}

This argument is well supported by our line shape model calculation,
shown in Fig.\ \ref{fig: TiTe2 simulation}, using
the Green's function used in Ref.\ {[}\ref{Allen:1995:JPhysChemSol}{]}. In 
this calculation, the $\mathbf{k}_{F}$ angle is defined to be 16$^o$. Other
parameter values are in the vicinity of the values used in Ref.\ 
{[}\ref{Allen:1995:JPhysChemSol}{]}, and are chosen to describe the line shape
near the 16\( ^{o} \) spectrum shown in Fig.\ \ref{fig: TiTe2 data}.  For
illustration purpose, the electron band dispersion is taken to be linear.
The salient experimental features are reproduced well, i.e.\ the peaks and
their back-bending below \( \mu \) at 25K and their appearance across \(
\mu \) at 300K. Note that at 25K, \( \mu \) lies very close to the top of
the peak for \( \mathbf{k}\approx \mathbf{k}_{F} \). This happens because
the line shape near the FS crossing is a sharp peak very close to \( \mu
\), which is then pushed slightly to the left side of \( \mu \) by the
function \( f \). This can occur whether the line shape is FL or non-FL,
as long as there is a sharp peak near \( \mu \). We note that ARPES peaks
above \( \mu \) have been previously demonstrated beautifully for Ni
\cite{Greber:1997:PES-above-EF}.  However, the three dimensional nature of
Ni makes a line shape analysis much more difficult.

To be sure, there are differences between the data and the line shape
calculations.  First, the line shape calculated at 300K is too sharp --
after crossing it shows a distinct two-peak structure which is not
observed in the data. We attribute this to additional broadenings expected
at high temperature but not included in the modeling -- i.e.\ failure of our
approximation of a \( T \) independent \( A \)\@. As the result, the
theoretical simulation shows peaks dispersing above \( \mu \) farther up
in energy, while in the experiment the peak reaches a maximum energy at
14\( ^{o} \) and then bends back. Second, there are small differences in
various estimates of \textbf{k}\( _{F} \) that one can make based on the
line shapes. This is of great significance
because \( \mathbf{k}_{F} \) is one of the most basic quantities in
condensed matter physics. Therefore we discuss this matter in the rest of
this section.

\begin{figure}
  {\centering \resizebox*{0.95\columnwidth}{!}{\includegraphics{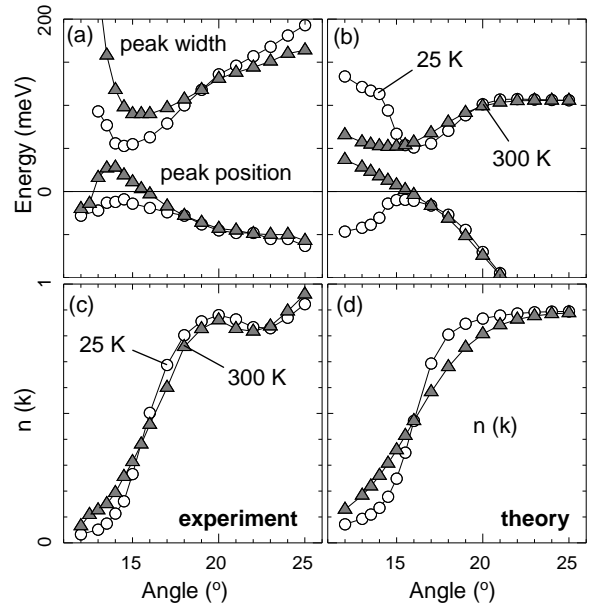}}
    \par}

\caption{Line shape attributes of experimental (Fig.\ \ref{fig: TiTe2
    data}) and theoretical (Fig.\ \ref{fig: TiTe2 simulation}) line shapes.
  \label{fig: TiTe2 line shape attributes}}
\end{figure}

We summarize in Fig.\ \ref{fig: TiTe2 line shape attributes} the peak
position, peak width, and area under the spectrum as a function of
\textbf{k}.  The various \textbf{k}\( _{F} \) estimates are summarized in
Table \ref{tab: TiTe2 various kF values}.  In the theory, the various
estimates are in quite good agreement with each other, if we ignore the
300 K peak width criterion which is expected to be rather unreliable due
to the neglect of the $T$ dependent broadening in $A$, but in the data,
they differ significantly. Notably, the minimum line width and the minimum
binding energy criteria applied to the 25 K data give significantly lower
values for \textbf{k}\( _{F} \).

\begin{table}
{\centering \begin{tabular}{|c|c|c|}
\hline 
Criterion&
 theory&
experiment\\
\hline 
\hline 
Peak position (300 K)&
16\( ^{o} \)&
16\( ^{o} \)\\
\hline 
Peak width (300 K)&
14.5\( ^{o} \)&
15.0--16.0\( ^{o} \)\\
\hline 
Peak position (25 K)&
16.5\( ^{o} \)&
14.5\( ^{o} \)\\
\hline 
Peak width (25 K)&
16\( ^{o} \)&
14.5\( ^{o} \)\\
\hline 
Fixed point of \( n(\mathbf{k}) \)&
16\( ^{o} \)&
15.5\( ^{o} \)\\
\hline 
\end{tabular}\par}

\caption{Various estimates of the \protect\( \mathbf{k}_{F}\protect \)
  angle for line shapes of Fig.\ \ref{fig: TiTe2 data} and 
  Fig.\ \ref{fig: TiTe2 simulation}. The true \protect\( \mathbf{k}_{F}\protect
  \) angle in the theory is 16\protect\( ^{o}\protect \), by definition.
\label{tab: TiTe2 various kF values}}
\end{table} 

The results in Table 1 are rather alarming. The two groups of criteria,
one being the minimum binding energy and minimum line width at 25 K and
the other
being the rest, are all good criteria in theory. However, applied to the
data, the two groups of criteria give quite different results. The
question is then which criterion is the most robust. We argue that this is
the peak-crossing criterion at 300K, because it relies on the single fact
that effectively the total spectral function \emph{\( A \)} is observed in
the \textbf{\( \mathbf{k},\omega \)} region of interest, due to the slow
variation of \( f \) in Eq.\ \ref{IkwT}. In contrast, the \textbf{k}\(
_{F} \) value extracted from data using the first group of criteria,
14.5\( ^{o} \), is clearly not good because at this angle the ARPES peak
exists above \( \mu \) at 300K, a fact very hard to understand if this
were indeed the crossing point.

In the least squares fit procedure applied to the 25 K data in Ref.\ 
{[}\ref{Allen:1995:JPhysChemSol}{]}, the \textbf{k}\( _{F} \) value was
determined, not surprisingly, to be 14.5\( ^{o} \).  Our conclusion is
then that this is \textit{not} a robust feature of the fit.  We have
demonstrated \cite{Gweon:1999:PhD-thesis} that the fit \textit{can} give
the correct \textbf{k}\( _{F} \) value if other effects are included in
the theory, such as electron hole asymmetry, \textbf{k}-mass, impurity
scattering, and the uncertainty in the chemical potential.  Recently, Kipp
\emph{et al.}\ \cite{Kipp:1999:PRL} presented a temperature
differential method to determine the \textbf{k}\( _{F} \) value for their
TiTe\( _{2} \) data, and determined the \textbf{k}\( _{F} \) angle to be
16.6\( ^{o} \). Their criterion is in principle equivalent to our \(
n(\mathbf{k}) \) fixed point criterion, which, as Table \ref{tab: TiTe2
  various kF values} shows, gives a result similar to that from our best
criterion. However, we note that the absolute value of their \textbf{k}\(
_{F} \) angle differs from ours by \( \sim 1^{o} \), which we interpret to
mean that the data are different. In addition, we have some cautionary
remarks about the \( n(\mathbf{k}) \) criterion. First, the argument of
Ref.\ {[}\ref{Kipp:1999:PRL}{]} depends critically on the assumption
of the electron-hole symmetry (within the energy range of \( \mu \pm
2k_{B}T \)).  Such symmetry may be more the exception than the rule.
Second, any \( T \) dependence in \( A \) will move the fixed point of \(
n(\mathbf{k}) \) away from \textbf{k}\( _{F} \).  In contrast, the
observation of the peak crossing \( \mu \) gives a more robust criterion
for \textbf{k}\( _{F} \). First, electron hole asymmetry gives an
asymmetric line shape but no change in peak position. Second, \( T \)
dependence in \( A \) broadens the line shape and gives a narrower range
of momentum over which the dispersing peak across \( \mu \) is observed,
as we infer from the comparison shown in Fig.\ \ref{fig: TiTe2 line shape
  attributes}.  However, as long as the dispersing peak is observed across
\( \mu \) for a finite range of \( \mathbf{k} \), the determination of \(
\mathbf{k}_{F} \) is not affected.

It is an interesting question how our results can be generalized to other
materials.  The condition \( 4k_{B}T\gg \hbar v^{\prime }_{F}\Delta
\mathbf{k} \) can be satisfied for either large \( T \), small \(
v^{\prime }_{F} \) or small \( \Delta \mathbf{k} \).  Note that the
current state of the art Scienta analyzer provides a \( \Delta \mathbf{k}
\) which is about a factor of 10 smaller than that used here, so that the
condition \( 4k_{B}T\gg \hbar v^{\prime }_{F}\Delta \mathbf{k} \) can be
easily satisfied for most materials at moderately high temperatures.
Therefore, it should be examined whether a behavior similar to that
reported here can be observed in other materials. If it turns out that
such behavior is not observed despite the condition \( 4k_{B}T\gg \hbar
v^{\prime }_{F}\Delta \mathbf{k} \), then that is a sign that the
intrinsic line shape is too broad or that the assumption of a dispersing
peak representing a metallic band is wrong.

\section{Li\protect\( _{0.9}\protect \)Mo\protect\( _{6}\protect \)O\protect\( _{17}\protect \) -- non-CDW non-FL Line Shape \label{sec: li bronze}}

\subsection{Background}

Li\( _{0.9} \)Mo\( _{6} \)O\( _{17} \) is a quasi-1D metal with two phase
transitions, at 24 K and 1.9 K\@. The transition at 24 K (\( T_{x} \)) is
not understood well, and that at 1.9 K is a superconducting transition.
The lowest temperature of our PES measurement is 12 K, so hereafter we
will concern ourselves with the phase transition at \( T_{x} \) only. This
transition shows up as a hump in the specific heat
\cite{Schlenker:1985:PhysicaBc} and a resistivity uprise
\cite{Greenblatt:1984:SolStComm}. However, no gap opening is observed in
the magnetic susceptibility \cite{Matsuda:1986:JPhys-C} and the optical
reflectivity \cite{Degiorgi:1988:PRB}. Furthermore, no structural
distortion is observed in X ray diffraction
\cite{Pouget:2000:libronze-no-Xray}. Therefore, we conclude that the
transition is not a CDW transition, because these measurements routinely
detect CDW gaps and Peierls lattice distortions in other materials such as
the blue bronze. Nevertheless, one may be tempted to explain the
resistivity uprise below \( T_{x} \) as a gap opening. Then a gap (2\(
\Delta \)) of 0.3 meV \cite{Greenblatt:1984:SolStComm} would be estimated.
Even if this gap picture is valid \cite{li-bronze-optics-gap-uncertainty},
such a gap is clearly not an ordinary CDW gap. Also, it should be noted
that there exists an explanation \cite{Matsuda:1986:JPhys-C} based on
localization physics.

\begin{figure}
  {\centering \resizebox*{0.9\columnwidth}{!}{\includegraphics{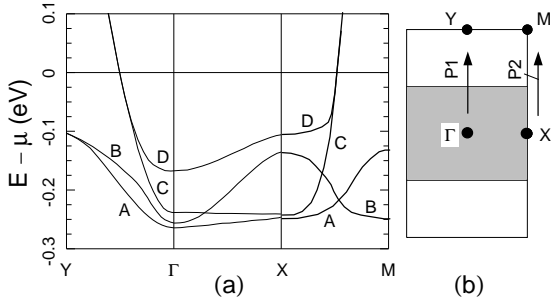}}
    \par}

\caption{(a) Band dispersions and (b) FS of Li$_{0.9}$Mo$_6$O$_{17}$. These
  results are taken from the band theory of Ref.\ 
  {[}\ref{Whangbo:1988:JAmChemSoc}{]}, except for the bands along the XM
  direction, which are sketches of the extension of the theory, based on
  our ARPES data. \label{fig: Li-band-theory}}
\end{figure}

The band structure of the Li purple bronze was calculated along the \(
\Gamma \)X and \( \Gamma \)Y directions by Whangbo \emph{et
  al}.\ \cite{Whangbo:1988:JAmChemSoc}, and is reproduced in Fig.\ \ref{fig:
  Li-band-theory}.  For completeness the figure also shows dispersions
along XM deduced from ARPES data presented further below.  According to
the calculation, there are four orbitally non-degenerate Mo \( 4d \) bands
near \( \mu \). The four bands are labeled as A,B,C and D in the order of
decreasing binding energy at the \( \Gamma \) point. Only C and D cross \(
\mu \) and they become degenerate before the crossing. The calculated FS
is perfectly 1D, and is given by \( k_{\Gamma Y}=0.45\Gamma Y \)\@. Note
that each of the A,C and D bands show large and similar dispersions along
the \( \Gamma \)Y direction and along the XM direction, while showing only
minor dispersions along the \( \Gamma \)X direction. On the other hand,
the band B shows similar dispersions along the directions \( \Gamma \)X 
and \( \Gamma \)Y, and shows opposite dispersions along
the directions \( \Gamma \)X and XM, parallel to the 1D chain.

So far, three groups have reported ARPES data on the Li purple bronze.
Initial data taken by Smith \emph{et al}.\ \cite{Smith:1993:PRL} and
our group \cite{Gweon:1996:JPhys-CondMatt} show almost dispersion-less peaks
with \( \mu \) crossings implied only by a spectral weight change.
Subsequent data taken by Grioni \emph{et al}.\ \cite{Grioni:1996:PhysScr}
with improved energy resolution, 15 meV, showed a hint of states crossing
\( \mu \), but these authors could not identify any crossing because their
angle sampling was coarse and because the dispersing peak intensity was
very weak. Our recent study \cite{Denlinger:1999:PRL} made use of a
geometry in which the bands C and D are strongest along a special \(
\mathbf{k} \) path, P2 of Fig.\ \ref{fig: Li-band-theory}, and discussed
the observed non-FL line shape. We will summarize this result in the next
section. Shortly after, Xue \emph{et al}.\ \cite{Xue:1999:PRL}
reported their new result obtained with a Scienta SES 200 analyzer, the
observation of a Fermi edge in $\mathbf{k}$-summed ARPES spectra above \(
T_{x} \), implying FL line shapes, and also a gap opening below \( T_{x}
\).  The gap below \( T_{x} \) was cited as being consistent with the
resistivity measurement, discussed above. Their finding of a FL line
shape, in conflict with not only our data \cite{Denlinger:1999:PRL}
but also \emph{all} the preceding data, was then attributed to the high
angular resolution of the new spectrometer. In our recent Comment
\cite{Gweon:2000:Comment} (also, see the Reply \cite{Smith:2000:Reply}),
we pointed out that (i) the basic band structure implied by their data is
in conflict with the band structure that emerges coherently from band
theory, our data and that of Grioni \emph{et al}.\ 
\cite{Grioni:1996:PhysScr}, (ii) our newly acquired similarly high
resolution data show the band structure same as the latter 
and continue to show non-FL
line shapes, (iii) their differing conclusion of a Fermi edge does not
flow from higher angle resolution, but rather from the fundamental
difference in the data, and (iv) their reported gap (80 meV) immensely
exceeds the value 0.3 meV implied in a gap model of the resistivity. In
this section, we give a more complete summary of our data than was
possible in previous publications, and also recapitulate some essential
findings of our published works. More details to support the points of
our Comment can be found in Sections \ref{sec: Li no gap}, \ref{sec: Li
  bands} and \ref{sec: Li discussion}, for points (iv), (i and ii) and (ii 
and iii) respectively.

\subsection{Absence of PES Gap Opening \label{sec: Li no gap}}

\begin{figure}
  {\centering
    \resizebox*{0.95\columnwidth}{!}{\includegraphics{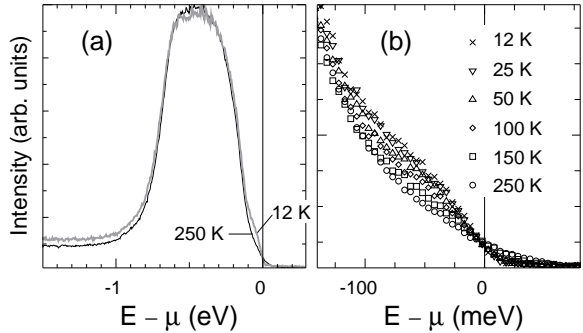}} \par}

\caption{\protect\( T\protect \) dependent angle integrated PES data for
  Li$_{0.9}$Mo$_6$O$_{17}$, measured at \protect\( h\nu =33\protect \) eV,
  \protect\( \Delta E=30\protect \) meV, \protect\( \Delta \theta =\pm
  6^{o}\protect \).
  \label{fig: Li-Tdep}}
\end{figure}

Perhaps the single most important feature of the Li purple bronze is that
it is, up to now, unique as a non-CDW quasi-1D metal studied by ARPES\@.
In this regard \( T \) dependent PES is of great interest. Fig.\ \ref{fig:
  Li-Tdep} shows our result, which does not show any sign of a gap
opening, within the energy resolution. The only observable change in the
line shape is the sharpening of the leading edge scaling with temperature
\cite{edge-sharpening-limit}. Our finding here is consistent with other
measurements discussed above, and does not necessarily preclude the
possibility that there is a small non-CDW gap.  We expect that such a gap,
if it exists, would have a value similar to or less than the value 0.3 meV
implied in a gap model of the resistivity.

\subsection{ARPES Line Shape -- Comparison with LL \label{sec: Li LL}}

The Li purple bronze is a good candidate for being a LL at \( T\gg T_{x}
\), because its electronic structure is highly 1D
\cite{Denlinger:1999:PRL} and is free of CDW formation.  Gapping
associated with the phase transition at \( T_{x}\), if it occurs at all,
is on such a low energy scale that a simple pseudo-gap effect cannot
underlie the NFL properties observed above \( T_{x} \).

\subsubsection*{LL Theory \label{sec: Li LL theory}}

A LL is defined as a system whose low energy fixed point is given by the
Tomonaga-Luttinger (TL) model
\cite{Tomonaga:1950:TL-model,Luttinger:1963:TL-model}. In this model,
electrons obey a linear band dispersion relation and the electron-electron
interaction is truncated so that electrons undergo only forward
scattering.  An amazing feature of this model is that it is exactly
solvable. The solution is characterized by two key features which
distinguish a LL from a 3D FL system: (1) an anomalous dimension \( \alpha
\) and (2) spin-charge separation. The first directly implies the absence
of Landau quasi-particles and the second means that the spin and charge
quantum numbers of an electron are carried by distinct elementary
excitations, i.e.\ waves of the spin density and the charge density.

PES line shapes for the TL model at \( T=0 \) are known in detail
\cite{Meden:1992:PRB,Voit:1993:PRB}.  The angle integrated
spectrum vanishes as a power-law \( |\omega |^{\alpha } \) at \( E_{F}
\)\@. ARPES spectra for \( \mathbf{k} \) inside the FS have two peaks (or
one peak and an edge if \( \alpha >0.5 \)) at positions corresponding to
charge and spin wave energies at \( \mathbf{k} \), and an edge for \(
\mathbf{k} \) outside the FS\@. We use the TL model
\cite{Meden:1992:PRB} obtained by truncating the general
electron-electron interaction of a continuum band to forward scattering
only. In this TL model with repulsive spin-independent interactions, the
spin velocity \( v_{s} \) is equal to \( v_{F} \), the charge velocity \(
v_{c} \) exceeds \( v_{F} \) by a factor \( \beta \) that is determined
entirely by \( \alpha \), and the edge singularity for \( \mathbf{k} \)
outside the FS disperses with velocity \( v_{c} \). We note that $v_s =
v_F$ is a property of a spin-rotationally invariant interaction in this TL
model, but not in the most general form of the TL model
\cite{Voit:1993:PRB}. For example, in the 1D Hubbard model which
is spin-rotationally invariant and can be mapped onto the TL model
in the low energy scale \cite{Schulz:1991:Hubbard-to-LL}, $v_s$ is
strongly renormalized.  Similarly, the relation between $\alpha$ and
$\beta$ of Ref.\ [\ref{Meden:1992:PRB}] is particular to the TL model
used there.  However, because the electronic states giving rise to the
quasi-1D properties of the Li purple bronze are based on an extended Mo
$4d$ wave-function, we believe that the TL model of Ref.\
[\ref{Meden:1992:PRB}] is the most appropriate starting point.

Solutions of the TL model can be extended to include weak interchain
couplings \cite{Kopietz:1997:PRB} and finite temperature
\cite{Nakamura:1997:finite-T-TL,Orgad:2000:finite-T-TL-and-LE}.  These
calculations show that the low energy LL behavior is modified within the
energy scales of the interchain hopping parameter \( t_{\perp } \) and
temperature \( T \), but the theory remains essentially the same as that
of the \( T=0 \) purely 1D model for energies larger than these.
Therefore, it is clear that in order to compare theory with experiment,
one needs to be aware of these energy scales and additionally a purely
experimental energy scale -- the energy resolution $\Delta E$. 
With this in mind, we
will use the \( T=0 \) solution of the TL model and include the \( T \)
dependence of the ARPES line shape only through the multiplication of the
\( f \) function in Eq.\ \ref{IkwT}. Note that the band theory and our \(
\mu \) intensity map \cite{Denlinger:1999:PRL} indicate \( t_{\perp
  }\approx 0 \) for the Li purple bronze, i.e.\ the FS is nearly flat as
predicted by band theory (Fig.\ \ref{fig: Li-band-theory}).

\subsubsection*{Comparison \label{sec: Li LL comparison}}

\begin{figure}
  {\centering \resizebox*{0.7\columnwidth}{!}{\includegraphics{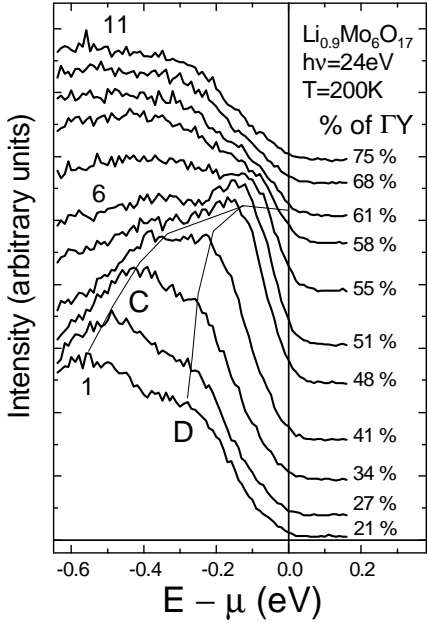}}
    \par}

\caption{ARPES data of Li$_{0.9}$Mo$_6$O$_{17}$ along the path P2 at
  \protect\( h\nu =24\protect \) eV, with \protect\( \Delta E=50\protect
  \) meV and \protect\( \Delta 
  \theta =\pm 1^{o}\protect \)\@.
\label{fig: Li-PRL-dispersion}}
\end{figure}

We show our ARPES data \cite{Denlinger:1999:PRL} taken along the
special path P2 in Fig.\ \ref{fig: Li-PRL-dispersion}. So far, this data
set remains as the one which shows the dispersing line shapes of both the
\( \mu \)-crossing C,D excitations most strongly.  This path was chosen to
intersect a spot in our \( \mu \) intensity map
\cite{Denlinger:1999:PRL} that is exceptionally bright for the photon
energy used and for the particular ARPES geometry of the Ames/Montana
end-station.

\begin{figure}
  {\centering
    \resizebox*{0.85\columnwidth}{!}{\includegraphics{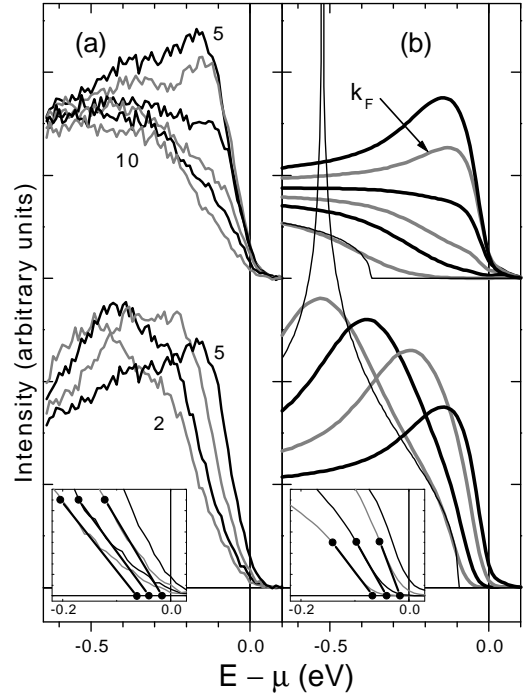}} \par}

\caption{(a) A replot of the ARPES data of Li$_{0.9}$Mo$_6$O$_{17}$ in
  Fig.\ \ref{fig: Li-PRL-dispersion} and 
  (b) LL theory simulation with \protect\( \alpha =0.9\protect \).
  \label{fig: Li-PRL-line-shape-comparison}}
\end{figure} 

Fig.\ \ref{fig: Li-PRL-line-shape-comparison} shows our comparison of the
ARPES data of Fig.\ \ref{fig: Li-PRL-dispersion} with line shapes for a
spin-independent repulsive TL model \cite{Meden:1992:PRB}. In the
absence of any LL line shape theory including interactions between two
bands, we apply line shapes calculated for the one-band TL model to the
two degenerate bands crossing $\mu$. 
The value
used for the anomalous dimension \( \alpha \) was 0.9.  The
electron-electron potential screening length \( r_{c} \) was taken to be
0.1 \AA\ so that the calculated spectral functions are well
within the validity limit of the universal LL behavior
\cite{Schonhammer:1993:PRB}.  We chose the \( v_{F} \) value so
that for this value of \( \alpha \) the renormalized charge velocity \(
\beta \)\( v_{F} \) ($\beta = 5$ for $\alpha=0.9$) coincides with the
linear dispersion that is observed
over a range to $\approx 0.2$ eV below \( \mu \) for peaks C and D while
they are degenerate and to $\approx 0.5$ eV for peak C alone. Thin lines
in Fig.\ \ref{fig: Li-PRL-line-shape-comparison} (b) show \(
A(\mathbf{k},\omega ) \)
without any experimental broadening, and demonstrate the charge peak and
the spin edge for \( \mathbf{k} \) inside the FS and the charge edge for
\textbf{\( \mathbf{k} \)} outside the FS\@.  The theoretical simulation
gives the \( \mu \) weight close to that of the data, that being the
reason for choosing an \( \alpha \) value a little larger than the value
0.6 deduced from the power law exponent for the angle integrated spectrum
\cite{Gweon:1999:PhD-thesis}.  Similarities of the theory and the data
also include the velocity \( v_{s} \) = \( v_{F} \) of the leading spin
edge
movement before crossing (see insets), significant because the agreement
is \emph{not forced} by our procedure for choosing parameter values, and
the loss of a spectral peaky upturn after the crossing. The general
goodness of the agreement for spectra 5 through 7 leads us to take the
value $k_{\Gamma Y} = 51$\% of $\Gamma Y$ as our best estimate for
$\mathbf{k}_F$. Differences between the data and the theory include there
being more intensity in the gap region before crossing and there being a
much slower backward movement of the charge edge after crossing, which we
attributed to the effects of 3D kinematics and to the presence of other
bands not included in the model, respectively.

\subsection{ARPES and Band Theory \label{sec: Li bands}}

\begin{figure}
  {\centering
    \resizebox*{0.95\columnwidth}{!}{\includegraphics{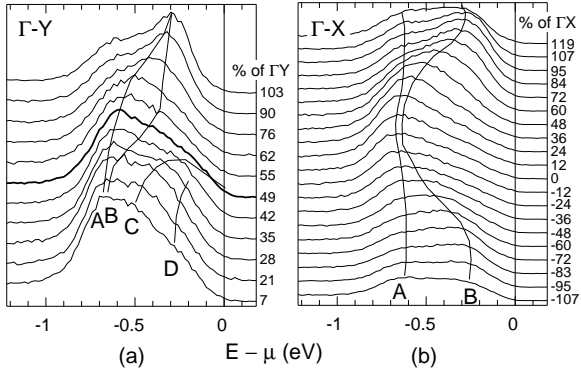}} \par}

\caption{ARPES data of Li$_{0.9}$Mo$_6$O$_{17}$
  taken along \protect\( \Gamma \protect \)Y (\protect\(
  T=200\protect \) K) and \protect\( \Gamma \protect \)X (\protect\(
  T=50\protect \) K), at \protect\( h\nu =24\protect \) eV, \protect\(
  \Delta E=100\protect \) meV, \protect\( \Delta \theta =\pm 1^{o}\protect
  \). Thick line corresponds to $\mathbf{k} = \mathbf{k}_F$ spectrum. 
\label{fig: Li-GXandGY-lowRes}}
\end{figure}

It is equally as important to know the global band structure as it is to
know what happens near the \( \mu \) crossing. Fig.\ \ref{fig:
  Li-GXandGY-lowRes} shows ARPES data along \( \Gamma \)X and \( \Gamma
\)Y\@. The sample surface is literally the same as the surface used in
Fig.\ \ref{fig: Li-PRL-dispersion}. Here, the most easily observed
features are the A,B bands. Along \( \Gamma \)Y, band C is observed to
cross \( \mu \) and a hint of D is seen near the \( \Gamma \) point. The
data along \( \Gamma \)X show the A,B bands most clearly. Note also that
there is an unexplained tendency, seen also \cite{Gweon:1996:JPhys-CondMatt} in
other compositionally similar bronzes KMo\( _{6} \)O\( _{17} \) and NaMo\(
_{6} \)O\( _{17} \), for non-dispersive weight to cling to the bottom of
the band. 

\begin{figure}
  {\centering
    \resizebox*{0.95\columnwidth}{!}{\includegraphics{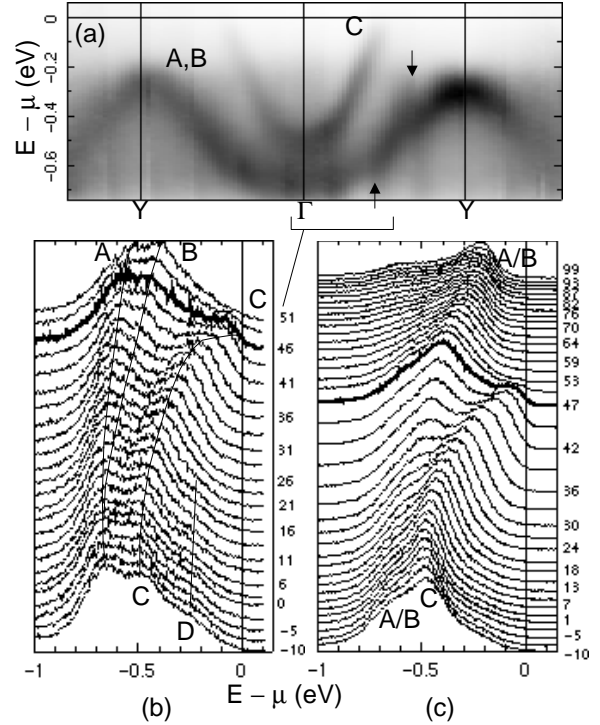}} \par}

\caption{(a), (b) ARPES data of Li$_{0.9}$Mo$_6$O$_{17}$
  taken along \protect\( \Gamma \protect \)Y at
  \protect\( h\nu =24\protect \) eV, \protect\( T=250\protect \) K,
  \protect\( \Delta E=35\protect \) meV, \protect\( \Delta \theta =\pm
  0.18^{o}\protect \) and (c) at \protect\( h\nu =30\protect \) eV,
  \protect\( T=200\protect \) K, \protect\( \Delta E=36\protect \) meV,
  \protect\( \Delta \theta =\pm 0.18^{o}\protect \)\@. For the meaning of
  arrows in (a), see text. Thick lines in (b) and (c) are $\mathbf{k} =
  \mathbf{k}_F$  spectra.
  \label{fig: Li-GY-Scienta}}
\end{figure}

The data taken along \( \Gamma \)Y with the Scienta analyzer are shown in
Fig.\ \ref{fig: Li-GY-Scienta}.  The gray scale map, shown in (a), spans
more than 1.5 unit cells in \textbf{\( \mathbf{k} \)} space. This map was
obtained by merging 12 overlapping angular mode Scienta data, and
therefore the intensity profiles in the overlapping angle (i.e.\ momentum)
regions do not always connect perfectly smoothly (e.g., see regions
pointed by arrows) due to the discontinuous change of the ARPES geometry
from one angular scan to the next.  Nevertheless, it is clear that the map
shows the C band dispersing to \( \mu \), and that the dispersions of the
A,B bands are consistent with the bulk crystal periodicity. EDC's are
shown in (b) and (c). Here, due to the better angle and energy
resolutions, the bands are resolved better than in Fig.\ \ref{fig:
  Li-GXandGY-lowRes}.  In particular, the C band is clearly observed to
cross \( \mu \) in (b) and (c).  In (b), the splitting of the A,B bands is
clearly observed. The small difference between the \( \mathbf{k}_{F} \)
value for these data and that for the preceding ARPES data is attributed
to a small variation of the Li ion numbers on the surface for different
cleaved surfaces. Note also that the line shape at \( \mathbf{k}_{F} \)
shows an increased \( \mu \) weight relative to the peak height. We
attribute this to the improved angle resolution, which we will discuss
shortly.

\begin{figure}
  {\centering
    \resizebox*{0.95\columnwidth}{!}{\includegraphics{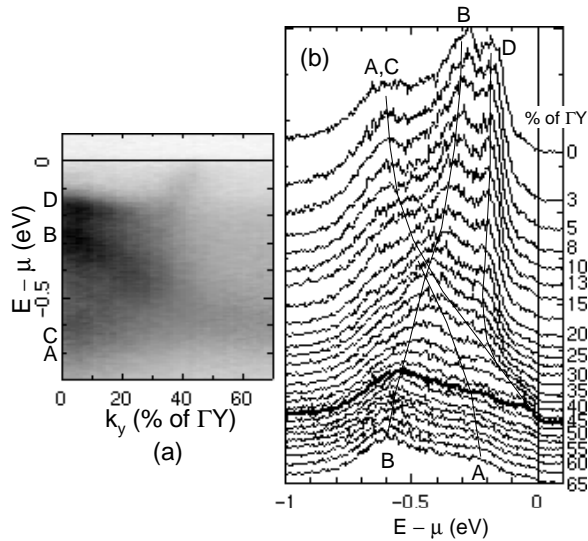}} \par}

\caption{Scienta data of Li$_{0.9}$Mo$_6$O$_{17}$ 
  taken along the special direction P2 
  (Fig.\ \ref{fig: Li-band-theory}) at \protect\( h\nu =24\protect \) eV,
  \protect\( T=30\protect \) K, \protect\( \Delta E=26\protect \) meV and
  \protect\( \Delta \theta =\pm 0.18^{o}\protect \)\@. Thick line
  corresponds to $\mathbf{k} = \mathbf{k}_F$ spectrum.
  \label{fig: Li-special-Scienta}}
\end{figure}

The experimental bands observed along \( \Gamma \)Y and \( \Gamma \)X are
in good agreement with the band theory of Fig.\ \ref{fig: Li-band-theory},
except for the overall band width. We now show Scienta data taken along
the special path P2 in Fig.\ \ref{fig: Li-special-Scienta}. Compared to
the data of Fig.\ \ref{fig: Li-PRL-dispersion}, the peaks are clearly
better resolved. Especially, the bands B and D are completely resolved,
much like in the band theory. As a result, the dispersion of peak B, which
could not be observed in the data of Fig.\ \ref{fig: Li-PRL-dispersion},
is now clearly observed, and the dispersion of peak D is now clearer. Note
that the relative intensities of the peaks in this data set are not
identical with those of Fig.\ \ref{fig: Li-PRL-dispersion}. The difference
is due to the difference in the geometry of the two experiments, i.e.\ a
different photon incidence angle relative to the surface normal. In
particular, the peaks C,D are not as strong as in Fig.\ \ref{fig:
  Li-PRL-dispersion}. Even so, the line shape at the \( \mu \) crossing
point is clearly observed and shows much more weight than that of the line
shape of Fig.\ \ref{fig: Li-PRL-dispersion}.  We discuss this aspect of
the data next.

\subsection{Discussion \label{sec: Li discussion}}

In the last section, we showed that the $(\mathbf{k}_F,\mu)$ weight
increases as the angle resolution is improved. This is an expected general
behavior if the $\mathbf{k} = \mathbf{k}_F$ spectrum is strongly peaky at
$\mu$. For example, the TL line shape for \( \mathbf{k}_{F} \) has a power
law behavior \( |\omega |^{\alpha -1} \), \emph{diverging} at \( \mu \)
for $\alpha < 1$. Therefore, as is easily shown by direct numerical
simulation, when the angle resolution is improved gradually, the \( \mu \)
weight steadily increases.  Thus the mere observation of more \( \mu \)
weight with better angle resolution is not by itself evidence for a FL
line shape. In fact, with regard to the weight at $\mu$, the only sure way
of distinguishing the FL and non-FL
line shapes is to examine the angle integrated spectrum.

\begin{figure}
  {\centering \resizebox*{1\columnwidth}{!}{\includegraphics{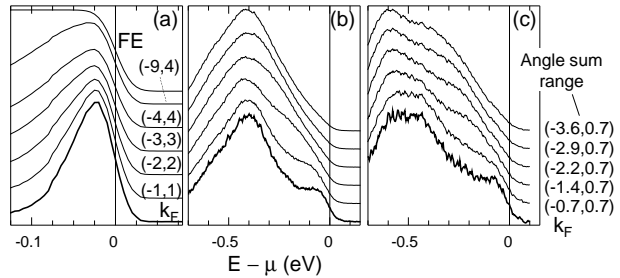}}
    \par}

\caption{Progressive angle integration of (a) TiTe\protect\( _{2}\protect
  \) data (Fig.\ \ref{fig: TiTe2 data} (a)), (b) Li$_{0.9}$Mo$_6$O$_{17}$ data
  (Fig.\ \ref{fig: Li-GY-Scienta} (c)), and (c) Li$_{0.9}$Mo$_6$O$_{17}$ data
  (Fig.\ \ref{fig: Li-GY-Scienta} (b)). The negative (positive) number in
  the angle sum range means the angle for which the band energy \protect\(
  \epsilon
  (\mathbf{k})\protect \) is negative (positive). In (a), a Fermi edge
  (FE) is drawn at the top. (b) and (c) have the same formats.  The stated
  angle sum range does \emph{not} include the inherent angle resolution of 
  an individual spectrum, $\pm 1^o$ for (a) and $\pm 0.18^o$ for (b) and (c).
  \label{fig: Li-progressive-angle-integration}}
\end{figure}

Theoretically the \( \mathbf{k} \)-sum (angle integration) of the FL line
shape and that of the TL line shape give quite different results.  The
former gives a Fermi edge and the latter gives a power law.  It is very
important to test whether this difference can be seen experimentally. Here
we examine the data from our FL reference TiTe\( _{2} \) and from the Li
purple bronze to see if this theoretical scenario comes true.  Fig.\ 
\ref{fig: Li-progressive-angle-integration} shows the result.  As more and
more angles are summed, the TiTe\( _{2} \) line shape converges to a Fermi
edge shape. The steep decrease on the high binding energy side of the
spectrum is due to the narrow band width of the Ti \( 3d \) band. In great
contrast, the line shape for the Li bronze loses the edge shape rapidly as
more and more angles are summed, converging to the power law behavior we
have observed using an angle integrating spectrometer in our home lab
\cite{Gweon:1999:PhD-thesis}. Our observation here directly contradicts
the claim of Xue \emph{et al.}\ \cite{Xue:1999:PRL} of a Fermi
edge in a partially angle integrated spectrum.

The Li purple bronze remains as a unique case of a CDW-free quasi-1D metal
accessible by photoemission. The global band structure is well understood
in both theory and experiment.  Furthermore, the two main characteristics
of the LL, the anomalous dimension $\alpha$ and the spin-charge
separation, are identified in the data, the former in the absence of Fermi
edge and the power-law onset of the angle integrated spectral function and
the latter by interpreting, appropriately for $\alpha > 0.5$, the
$\mu$-crossing peak as the charge peak and the extrapolated finite energy
onset of the peak as the spin edge. Among these features the spin edge is
perhaps the least convincing feature due to the smoothness of the edge in
the data.  A likely source of this smoothness is 3D kinematics, as we
mentioned at the end of Section \ref{sec: Li LL comparison}. Also, the
spin velocity itself needs to be understood better. Within the TL theory
used here, it is supposed to be the same as the band velocity, but the
value used in our line shape analysis is about a factor of 2 too small
compared with the value of the band theory (Fig.\ \ref{fig:
  Li-band-theory}). This may be due to the inaccuracy of the tight binding
calculation or the simplistic nature of the LL theory we used.  For
example, it is reasonable to think that the backward scattering terms,
completely neglected here, will in reality have a residual effect of
changing the relation between $\alpha$ and the spin and/or charge
velocity, analogously to the Hubbard model case mentioned in Section
\ref{sec: Li LL theory}. Therefore, a more detailed understanding calls
for a first principles band theory and a better understanding of the
electron-electron interactions.

\section{Blue Bronze -- CDW non-FL Line Shape \label{sec: blue bronze}}

\subsection{CDW Theory}

Because the CDW is an essential ingredient for describing the physics of
the blue bronze, we start this section with a discussion of CDW theory.
The so-called FS nesting condition, that one part of the FS matches
another part of the FS via a translation by a single wavevector, implies
an instability towards the formation of a CDW with periodicity given by
the nesting vector and the consequent formation of a CDW gap at \( \mu \)
\cite{Peierls:1955:quantum-theory-of-solids-book}.  Via electron-phonon
coupling, the lattice is distorted with the same wave vector.  The nesting
condition is fulfilled perfectly in 1D, and can be met approximately but
with increasing difficulty in 2D and 3D.  In the standard mean-field
description of the Fr\"{o}lich Hamiltonian
\cite{Gruner:1994:density-wave-book}, which is formally equivalent to the
BCS theory of superconductivity, a phase transition occurs at a finite
temperature \( T_{c} \) where the lattice modulation at the nesting vector
becomes static. Below \( T_{c} \), the CDW gap opens up with the same \( T
\) dependence as the BCS gap. Above \( T_{c} \), the electronic state is
that of a simple FL metal.

The mean-field picture of the CDW requires significant modification to
account for fluctuations, especially in low dimensions.  In fact, for a
perfectly 1D system, a finite \( T \) phase transition does not occur, due
to the well known fact
\cite{Mermin:1966:mermin-theorem,Hohenberg:1967:no-phase-transition-in-low-D}
that in 1D the entropy increase associated with the break-up of long range
order into many short range orders wins over the energy minimization
associated with the long range order. Therefore, it is obvious that a CDW
fluctuation theory must be also a 3D theory in order to have the power to
predict a realistic finite \( T \) phase transition.

Much work has been done on CDW fluctuations. A study by McKenzie and
Wilkins \cite{McKenzie:1992:lattice-fluc-at-low-T} predicts a significant
filling-in of the gap region at low \( T \) due to CDW fluctuations and
also quantum lattice fluctuations.  Most theories are concerned however
with \( T \ge T_{c} \).  Some theories \cite{Sadovskii:1979,
  Millis:2000:cdw-pseudo-gap} treat fluctuations exactly but deal with a
single 1D chain, and others, e.g.\ that of Rice and Str\"assler (RS)
\cite{Rice:1973:RiceAndStraessler}, treat fluctuations perturbatively but
include interchain couplings. We find the latter type of theory to be more
useful because there are enough ingredients to permit a realistic
comparison to experiments.  We also find that the line shapes predicted
using the former type of theory, specifically that of Sadovskii
\cite{Sadovskii:1979}, are actually quite similar to those predicted by
the RS theory for the same correlation length.

The result of the RS theory is summarized in its self energy, \( \Sigma
_{RS}(\mathbf{k},\omega ) \),
\begin{equation}
\label{RS-Sigma}
\frac{\psi }{-i\, f}\, \log \left( 1-\frac{i\, f}{\omega /\psi -\epsilon (\mathbf{k}\pm \mathbf{q}_{CDW})/\psi -i\, \gamma }\right) 
\end{equation}
where \( \psi \), \( f \) and \( \gamma \) are all \( T \) dependent
quantities. \( \psi \) is the ``pseudo-gap'' parameter, i.e.\ the
root-mean-square fluctuation of the order parameter, and \( \gamma =\hbar
v_{F}/(\xi \psi ) \) where \( \xi \) is the correlation length of the CDW
fluctuation. The parameter \( f \) is basically an effective 3D coupling
strength parameter, and distinguishes this theory from that of Lee, Rice
and Anderson (LRA) \cite{Lee:1973:LRA}, which is a perturbative 1D theory.
RS theory predicts, in the limit of strong fluctuation (i.e.\ \emph{\(
  f\leq O(1) \)}),
\begin{equation}
\label{RS-Tc}
T_{c}/T_{MF}=0.26f(T_{c})^{1/3},
\end{equation}
which implies \( T_{c}\ll T_{MF} \)\@.

\begin{figure}
  {\centering \resizebox*{0.4\columnwidth}{!}{\includegraphics{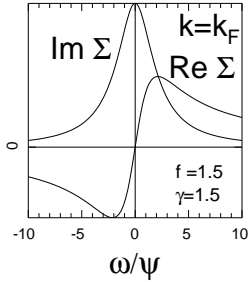}}
    \par}

\caption{Self energy for the RS theory. Theoretical parameters are those
  for our simulation of blue bronze data at 250 K\@.
  \label{fig: RS theory Sigma}}
\end{figure}

The self energy of Eq.\ \ref{RS-Sigma} is shown in Fig.\ \ref{fig: RS
  theory Sigma}.  In contrast to the situation in the simple mean-field
solution, the normal state is not a FL in this theory. Im\( \Sigma \) has
a finite value and a negative curvature at \( \mu \), and Re\( \Sigma \)
has a positive slope at \( \mu \), directly contradicting the well-known
FL self energy behavior \cite{Luttinger:1961:wsquare}.  These properties
were pointed out by McKenzie \cite{McKenzie:1996:non-FL-line-shape} for
Sodovskii's results.

\subsection{Background}

K\( _{0.3} \)MoO\( _{3} \), is one of the most intensely studied CDW
materials, and yet some basic properties are still difficult to
understand.  Its CDW wavevector, studied by X ray diffraction
\cite{Pouget:1983:lattice-modulation} and neutron scattering
\cite{Escriberfilippini:1985:LecNoPhys}, is unusual in that it shows a \( T
\) dependence. The magnetic susceptibility and the resistivity
measurements are intriguing because in the normal state, the magnetic
susceptibility increases steadily up to the highest measured temperature of
720K \cite{Johnston:1984:bb-spin-gap} while the resistivity shows
perfectly metallic behavior. Later in this section we will discuss these
issues further.

The crystal structure of the blue bronze is centered monoclinic
\cite{Graham:1966:bb-struct,Ghedira:1985:bb-struct}, and the repeating
motif is the Mo\( _{10} \)O\( _{32} \) chain which defines the \emph{\( b
  \)} axis -- the ``easy'' axis. The quasi-1D nature of the electron
conduction is shown by the resistivity \cite{Brusetti:1981} and the
optical properties \cite{Travaglini:1981:bb-optics}. The basic band
structure was calculated first by Whangbo and Schneemeyer
\cite{Whangbo:1986:InorgChem}. The calculation shows that two orbitally
non-degenerate Mo \emph{\( 4d \)} bands are partially occupied by
electrons donated by the K\( ^{+} \) ions, making the material conducting.
In the notation of these authors, which we follow here, the BZ boundary
along the quasi-1D \emph{b} axis is called the X point, and we wish to
remind readers that the equivalent point for the Li purple bronze was
called the Y point in Section \ref{sec: li bronze}.

Veuillen \emph{et al.}\ \cite{Veuillen:1987:EurophysLett} first reported ARPES
results on the blue bronze. Their result shows a single broad peak
dispersing to a \( \mu \) crossing at a \textbf{k} value in good agreement
with the CDW wavevector. A subsequent high resolution angle integrated PES
study by Dardel \emph{et al.}\ \cite{Dardel:1991:PRL} showed a
spectrum with anomalously low \( \mu \) weight and no distinct Fermi edge.
They also reported \( T \) dependent data \cite{Dardel:1992:EurophysLett}
which showed evidence of a CDW gap opening. Breuer \emph{et al.}\ 
\cite{Breuer:1994:bb-surface-damage} did a detailed study of the surface
damage caused by photon and electron/ion bombardment. The major symptom of
surface damage is the emergence of a peak at \( \sim 2 \) eV binding
energy and the shift of the spectral weight at \( \mu \) to higher binding
energy. By taking precautions to minimize photon bombardment above the
absorption threshold energy (\( >36 \) eV), we have obtained ARPES spectra
\cite{Gweon:1996:JPhys-CondMatt,Claessen:1995:JElecSpecRelPh} with strikingly low
inelastic background and two dispersing peaks crossing \( \mu \). We have
also demonstrated \cite{Gweon:1996:JPhys-CondMatt} that the two dispersing peaks
cross \( \mu \) at different \textbf{k} values by taking a \( \mu \)
intensity map.  The two dispersing peaks were subsequently reproduced by
Grioni \emph{et al.}\ \cite{Grioni:1996:PhysScr}, and more recently by
Fedorov \emph{et al.}\ \cite{Fedorov:2000:JPhys-CondMatt}. The latter authors
also reported a \( T \) dependence of the nesting vector measured in ARPES
along \( \Gamma \)X to be in good agreement with that of the CDW wave
vector measured in neutron scattering (See, however, our
discussion in Section \ref{sec: blue bronze line shape discussion}).

The non-FL line shape of the blue bronze, namely the absence of a PES
Fermi edge, remains unexplained. In this section, we show that this
feature is not reconcilable with the standard CDW picture even when the
pseudo-gap mechanism is included.

\subsection{Data \label{sec: blue bronze data}}

In this section we introduce some new ARPES data, as well as summarize
some key results from the literature. In particular new \( T \) dependent
high resolution data are, to our knowledge, the first to show how the
ARPES line shapes of the EDC's change across
\( T_C \) (180 K)\@.

\begin{figure}
  {\centering \resizebox*{0.6\columnwidth}{!}{\includegraphics{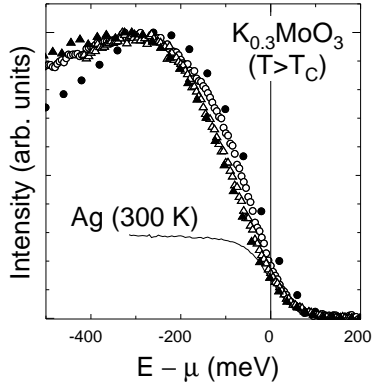}}
    \par}

\caption{Angle integrated photoemission data of K$_{0.3}$MoO$_3$ in the
  normal metallic state. Filled circles: angle sum of data in
  Ref.\ {[}\ref{Claessen:1995:JElecSpecRelPh}{]} (also shown in Fig.\ 
  \ref{fig: blue bronze ARPES 20eV}), \protect\( h\nu =20\protect \) eV,
  \protect\( \Delta E=100\protect \) meV, \protect\( \Delta \theta =\pm
  10^{o}\protect \), \protect\( T=220\protect \) K\@. Empty circles: this
  work, \protect\( h\nu =13\protect \) eV, \protect\( \Delta E=27\protect
  \) meV, \protect\( \Delta \theta =\pm 6^{o}\protect \), \protect\(
  T=250\protect \) K\@. Empty triangles and line:
  Ref.\ {[}\ref{Gweon:1996:JPhys-CondMatt}{]}, \protect\( h\nu =21.2\protect \)
  eV, \protect\( \Delta E=33\protect \) meV, \protect\( \Delta \theta =\pm
  12^{o}\protect \), \protect\( T=300\protect \) K\@. Filled triangles:
  Ref.\ {[}\ref{Dardel:1992:EurophysLett}{]}, \protect\( h\nu =21.2\protect \)
  eV, \protect\( \Delta E=20\protect \) meV, \protect\( \Delta \theta =\pm
  3^{o}\protect \), \protect\( T=313\protect \) K\@. 
  \label{fig: blue bronze angle integ. collection}}
\end{figure}

Fig.\ \ref{fig: blue bronze angle integ. collection} shows various high
resolution angle integrated photoemission spectra and contrasts them with
a reference spectrum taken on Ag. These angle integrated spectra are taken
in the metallic state well above the phase transition. Nevertheless, the
Fermi edge that is characteristic of a 3D metal is completely absent. It
is interesting to note that a related material, NaMo\( _{6} \)O\( _{17}
\), whose electronic structure is that of three weakly interacting 1D
chains oriented at 120 degrees to one another in planes, {\em does} show a
Fermi edge, as we reported in Ref.\ {[}\ref{Gweon:1996:JPhys-CondMatt}{]}.

\begin{figure}
  {\centering \resizebox*{0.95\columnwidth}{!}{\includegraphics{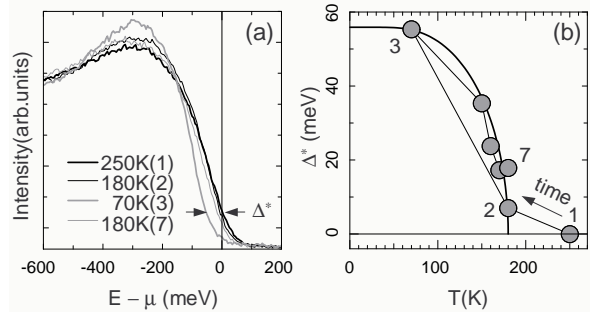}}
    \par}

\caption{(a) \protect\( T\protect \) dependent angle integrated
  photoemission data of K$_{0.3}$MoO$_3$. 
  \protect\( h\nu=13\protect \) eV, \protect\( \Delta
  E=27\protect \) meV, \protect\( \Delta \theta =\pm 6^{o}\protect \).
  (b) The ``gap'' parameter \protect\( \Delta ^{*}\protect \) deduced from
  data. The thick  curve is the BCS gap function. 
  \label{fig: blue bronze T dep. angle int.}}
\end{figure}

One may wonder whether the absence of the Fermi edge is merely the result
of a bad surface, not representative of the bulk. However, this is not so.
Our \( \mu \) intensity map showed a FS consisting of two pairs of lines
which imply a nesting vector in agreement with the CDW wavevector
\cite{Escriberfilippini:1985:LecNoPhys}.  Also the bulk phase transition is
clearly detectable in \( T \) dependent measurement of angle integrated
spectra \cite{Dardel:1992:EurophysLett}.  Fig.\ \ref{fig: blue bronze T dep.
  angle int.} shows our own \( T \) dependent PES data, taken with a
Scienta SES 200 analyzer using its angle integrated mode.  The sequence of
\( T \) in the measurement was 250 K, 180 K, 70 K, 150 K, 160 K, 170 K and
180 K. This \( T \) sequence was deliberately chosen to reveal any effects
of irreversible sample damage during the \( T \) variation.  The two 180 K
spectra in Fig.\ \ref{fig: blue bronze T dep. angle int.}(a), taken
initially and at the end of the cycle, are almost identical with each
other, indicating that the data are largely free from such an undesirable
irreversibility.
 
The spectral change observed in Fig.\ \ref{fig: blue bronze T dep. angle
  int.}  is in good agreement with the results reported in Ref.\ 
{[}\ref{Dardel:1992:EurophysLett}{]}.  Below the transition point (180 K), the
spectral edge moves to a higher binding energy, signaling gap opening.
These spectra should be contrasted with those of Fig.\ \ref{fig: Li-Tdep}
for the Li purple bronze, where a gap does not open.  To exactly quantify
the gap opening is a difficult task due to the odd line shape, and we
define a first approximation \( \Delta ^{*} \) (see Fig.\ \ref{fig: blue
  bronze T dep. angle int.} (a)) as the shift of the spectral edge at the
intensity value corresponding to \( \mu \) at 250 K\@.  Our method is
different from the one used by Dardel \emph{et al.}\ 
\cite{Dardel:1992:EurophysLett}, i.e.\ taking the inflection point to quantify
the gap opening, but gives similar results, shown in Fig.\ \ref{fig: blue
  bronze T dep. angle int.} (b).  The difference at 180 K of the final
point 7 from the initial point 2 is probably due to a slight degradation
of the surface.  Similar to the findings by Dardel \emph{et al.}\ 
\cite{Dardel:1992:EurophysLett}, the temperature variation of the gap opening
is roughly BCS like. The \( \Delta ^{*}(T=0) \) value deduced from our
procedure, 56 meV, is a lower bound, for reasons that we will discuss
later.

\begin{figure}
  {\centering \resizebox*{0.7\columnwidth}{!}{\includegraphics{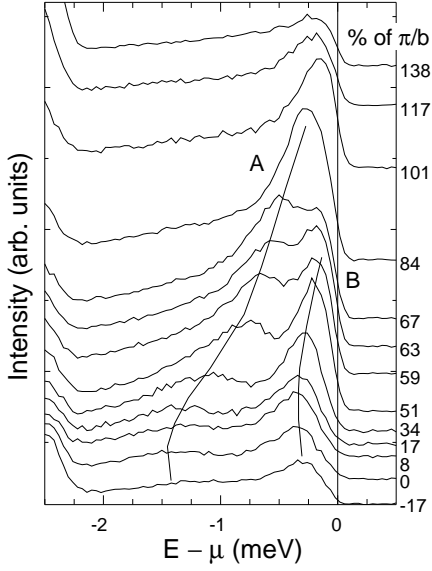}}
    \par}

\caption{ARPES data of K$_{0.3}$MoO$_3$ taken at \protect\( h\nu =20\protect \)
  eV, \protect\( \Delta E=100\protect \) meV, \protect\( \Delta \theta
  =\pm 1^{o}\protect \) and \protect\( T=220\protect \) K\@. Thin lines
  are guides to the eye for dispersions. 
  \label{fig: blue bronze ARPES 20eV} }
\end{figure}

Fig.\ \ref{fig: blue bronze ARPES 20eV} shows our early modest resolution
high temperature ARPES data, most of which were already reported in
Ref.\ {[}\ref{Claessen:1995:JElecSpecRelPh}{]}.  This ARPES data set
attests to a very clean surface because the background intensity level is
very low and there is no sign of the defect peak at \( 2 \) eV binding
energy. For this reason, we have included the angle sum of this data set
in the collection of angle integrated data of Fig.\ \ref{fig: blue bronze
  angle integ. collection}.

As we have shown before \cite{Gweon:1996:JPhys-CondMatt}, the two peaks of
Fig.\ \ref{fig: blue bronze ARPES 20eV} cross at distinct \textbf{k}\( _{F}
\) values. The \( \mu \) intensity map taken at \( h\nu =17 \) eV
\cite{Gweon:1996:JPhys-CondMatt} shows that band B crosses at \( \approx \)62 \%
of \( \pi \)/\emph{b}.  The exact crossing point of band A is not easily
determined because the map shows maximum intensity centered at the X
point. Similarly the EDC's of Fig.\ \ref{fig: blue bronze ARPES 20eV} and
Ref.\ {[}\ref{Gweon:1996:JPhys-CondMatt}{]} show an almost symmetric band having
a maximum at the point X.  Therefore, in these moderate resolution data,
we take the spectrum at the X point (e.g.\ the 101 \% spectrum in Fig.\
\ref{fig: blue bronze ARPES 20eV}) to be representative of the \textbf{k}
$=$ \textbf{k}\(_{F}\) spectrum.

\begin{figure}
  {\centering \resizebox*{1\columnwidth}{!}{\includegraphics{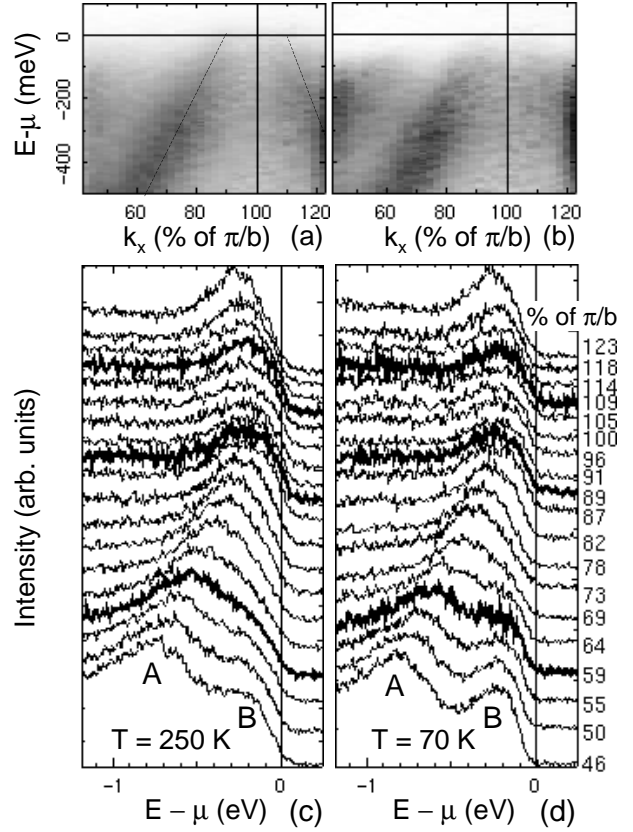}}
    \par}

\caption{\protect\( T\protect \) dependent ARPES data of K$_{0.3}$MoO$_3$
  taken at \protect\(
  h\nu =13\protect \) eV, \protect\( \Delta E=27\protect \) meV, and
  \protect\( \Delta \theta =\pm 0.18^{o}\protect \).  Thin lines in (a)
  are guides to the eye for dispersions. Thick lines in (c) and (d) are
  $\mathbf{k} = \mathbf{k}_F$ spectra.
  \label{fig: blue bronze T dep. ARPES}}
\end{figure}

\begin{figure}
  {\centering \resizebox*{0.95\columnwidth}{!}{\includegraphics{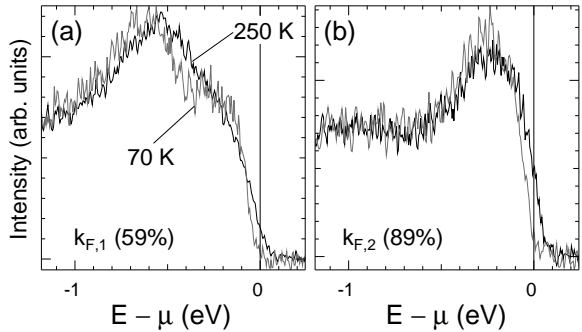}}
    \par}

\caption{\protect\( T\protect \) dependence of spectra at FS crossings for
  the K$_{0.3}$MoO$_3$ data of Fig.\ \ref{fig: blue bronze T dep. ARPES}. 
  \label{fig: blue bronze T dep. kF}}
\end{figure}

Fig.\ \ref{fig: blue bronze T dep. ARPES} shows our new high resolution
data taken with \( 13 \) eV photons near
the X point at 250 K and 70 K.  Each data set was taken immediately after
taking the angle integrated 250 K and the 70 K spectra of Fig.\ \ref{fig:
  blue bronze T dep. angle int.}, respectively, i.e., they were taken on
an undamaged surface showing the CDW gap opening.  This new data set is an
improvement over the previous lower resolution data set, in that the
crossing near the X point is now resolved due to better angle resolution.
The \( \mu \) intensity pattern shows a minimum at the X point, instead of
a maximum, and it enables identification of FS crossing points for band A
(90 \%) and for band B (59 \%). For later discussion in Sections \ref{sec:
  BB CDW line shape} and \ref{sec: conclusion}, we directly compare 
the \emph{T} dependence of the spectra at these crossing points in Fig.\
\ref{fig: blue bronze T dep. kF}. From these crossings, we get an estimate
of the CDW 
wavevector of 75 \% of \( b^{*} \), which seems to be in good agreement
with the observed value which varies from 72 \% to 75 \% as \( T \) varies
from 180 K to 0 K. We note that from our \( \mu \) intensity map
\cite{Gweon:1996:JPhys-CondMatt} one cannot rule out a somewhat 2D FS for band
A. In this case we would have an imperfect nesting condition such as we
have observed for SmTe\( _{3} \) \cite{Gweon:1998:PRL} where the
nesting vector along one particular direction (e.g.\ the \( \Gamma \)X
direction) generally differs from that of the CDW wavevector, which is a
compromise value determined by global energy minimization across the
entire 2D FS.
 
A surprising finding from comparing the data of Fig.\ \ref{fig: blue
  bronze ARPES 20eV} and Fig.\ \ref{fig: blue bronze T dep. ARPES} is that
the widths of the A,B peaks and also the $({\mathbf k}_F,\mu)$ weight
relative to the peak height do \emph{not} change significantly as the
angle resolution is improved. This is a direct spectroscopic contrast
between the Li purple bronze and the blue bronze.

\subsection{Line Shapes}

In this section, we compare CDW and LL line shape theories with our data.
The theories used here are single band theories, while there are actually
two $\mu$ crossing bands in the blue bronze. The $\mu$ crossing line
shapes of band B are obscured by the presence of band A, while those of
band A are isolated near the X point. Therefore, our ARPES comparison is
focused on band A\@.

\subsubsection*{CDW \label{sec: BB CDW line shape}}

\begin{figure}
  {\centering \resizebox*{0.95\columnwidth}{!}{\includegraphics{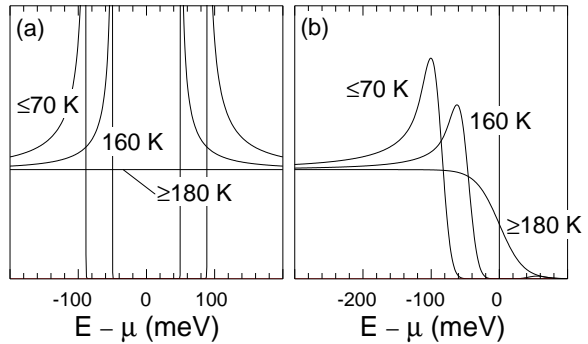}}
    \par}

\caption{(a) Temperature dependent density of states and (b) angle
  integrated photoemission data calculated within the mean-field CDW
  theory. Strictly speaking, theory is \protect\( T\protect \) dependent
  below 70 K, but is negligibly so. Experimental energy resolution for the
  data of Fig.\ \ref{fig: blue bronze T dep. angle int.}  is included in (b).
  \label{fig: CDW theory MF angle int.}}
\end{figure}

An obvious starting point for comparing the data to theory is the mean
field CDW theory.  The prediction of the mean field theory for the ARPES
line shape is simple: the band dispersion relation \( \epsilon
(\mathbf{k}) \) is replaced by \( -\sqrt{\epsilon (\mathbf{k})^{2}+\Delta
  ^{2}} \) \cite{different-vs} and the gap \( \Delta(T) \) has the BCS \(
T \) dependence.  The magnitude of \( \Delta (0) \) for the blue bronze
shows a significant variation in the literature: \( \sim 40 \) meV
(resistivity \cite{Brutting:1995:PRB}), \( \sim 50 \) meV (magnetic
susceptibility \cite{Johnston:1984:bb-spin-gap}) and \( \sim 90 \) meV
(optics \cite{Degiorgi:1995:PRB}). Hereafter, we will use the result
from the optical measurement, consistent with taking our estimate of 56
meV to be a lower bound for $\Delta (0)$, as explained below. The angle
integrated spectral functions calculated in the mean field theory are
shown in Fig.\ \ref{fig: CDW theory MF angle int.}.
 
The mean field CDW theory cannot adequately describe our photoemission
data.  First of all, the normal state angle integrated PES data do not
show a Fermi edge (Fig.\ \ref{fig: blue bronze angle integ. collection}),
in contrast to the calculation of Fig.\ \ref{fig: CDW theory MF angle
  int.}.  Second, the experimental line shape changes in the angle
integrated PES data due to the CDW gap opening (Fig.\ \ref{fig: blue
  bronze T dep. angle int.}) are difficult to understand.  These changes
include the low $T$ intensity pile-up occurring at a much larger energy \(
\sim 0.3 \) eV compared to the gap energy, as was noted already by Dardel
\emph{et al}.\ \cite{Dardel:1992:EurophysLett}, and the existence of the
significant sub-gap tail at 70 K\@.  Third, the peak shift by \( -\Delta
\) expected to occur for the $\mathbf{k}=\mathbf{k}_F$ ARPES data is not
observable in Fig.\ \ref{fig: blue bronze T dep. kF}.  Instead, only an
intensity redistribution within the ARPES peak seems to occur. This can be
contrasted to the case of the high temperature CDW material SmTe\( _{3} \)
\cite{Gweon:1998:PRL}, for which the dispersion relation \(
-\sqrt{\epsilon (\mathbf{k})^{2}+\Delta ^{2}} \) \cite{different-vs} is
clearly observed.

It is an obvious next step to test whether the inclusion of CDW
fluctuations improves the comparison of the data with the CDW theory.
Evidences for CDW fluctuations are ample.  Below \( T_{c} \), a strong
sub-gap tail is observed in optics \cite{Degiorgi:1995:PRB}, in
qualitative agreement with the theory by McKenzie and Wilkins
\cite{McKenzie:1992:lattice-fluc-at-low-T} and also with our observation
of a strong sub-gap tail existing at 70 K\@.  This is why we take our
estimate \( \Delta ^{*} \) to be a lower bound. In this article, our main
interest however is in the fluctuations in the normal state above \( T_{c}
\)\@. Evidences for fluctuations above \( T_{c} \) are the diffuse
scattering observed by X ray experiments
\cite{Pouget:1985:JPhys-Paris,Schlenker:1989:Mo-bronze-book}, and the large
value of \( 2\Delta (0)/(k_{B}T_{c}) \), 5--12, compared to the mean field
value 3.52.

Next we estimate parameters for the RS theory. For the estimate of \(
T_{MF} \), we use the mean-field relation \( 3.52k_{B}T_{MF}=2\Delta (0)
\), and get \( T_{MF}=590 \) K for \( \Delta (0)=90 \) meV\@. Then, from
Eq.\ \ref{RS-Tc}, we get \( f(T_{c})=1.5 \).  Then, the weak,
and unimportant, \( T \) dependence of \( f \) is included as outlined in 
Ref.\ {[}\ref{Rice:1973:RiceAndStraessler}{]}. For our
\( f(T_{c}) \) value, Eq.\ 9 of Ref.\ 
{[}\ref{Rice:1973:RiceAndStraessler}{]} gives \( \psi (T_{c})=54 \) meV\@.
The pseudo-gap is expected to decrease as \( T \) increases, and a
calculation \cite{McKenzie:1995:cdw-theory} does show such behavior. In
our modeling, we simply ignore this \( T \) dependence.  By so doing, we
are somewhat overestimating the pseudo-gap above \( T_{c} \).  Note that
our estimate that \( \psi (T_{c}) \) is roughly half of \( \Delta (0) \)
is in good agreement with estimates by others
\cite{Johnston:1984:bb-spin-gap,Brutting:1995:PRB}.  For \( \hbar
v_{F} \), we measure the peak dispersion in the ARPES data and get \( \sim
4.5 \) eV\AA\ for band A and \( \sim 3 \) eV\AA\ for band B\@. The nesting
occurs between these two bands, and therefore \( v_{F} \) to be used in
the expression for \( \gamma \) should be an ``average'' of these two
values. Instead, we simply use the value for band B and again slightly
overestimate the effect of the pseudo-gap. Lastly, for the correlation
length \( \xi (T) \), we use the result measured by X-ray diffraction
\cite{Pouget:1985:JPhys-Paris,Tc-difference}.

\begin{figure}
  {\centering
    \resizebox*{0.95\columnwidth}{!}{\includegraphics{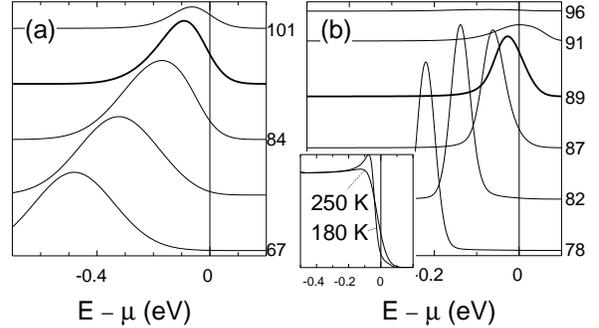}} \par}

\caption{Simulation of CDW fluctuation line shape (a) for the data of
  Fig.\ \ref{fig: blue bronze ARPES 20eV} and (b) for those of
  Fig.\ \ref{fig: blue bronze T dep. ARPES} (\protect\( T=250\protect \)
  K). Thick lines are \protect\( \mathbf{k}=\mathbf{k}_{F}\protect \)
  spectra.  Inset: Simulation of the data of Fig.\ 
  \ref{fig: blue bronze T dep. angle int.} at 250 K and 180 K\@. 
  \label{fig: blue bronze RS simulation}}
\end{figure}

Fig.\ \ref{fig: blue bronze RS simulation} shows our simulation. For the
angle integrated spectrum shown in the inset, note that the simulation
does show suppression of weight at \( \mu \). However, the \( \mu \)
weight at 250 K is significantly larger (\( 35 \) \%) than that for the 250 K 
data of Fig.\ \ref{fig: blue bronze T dep. angle int.}  (a) (25 \%).
Furthermore, the \( T \) dependence observed above \( T_{c} \) is far too
weak compared to the simulation. Perhaps most importantly, the simulation
shows a Fermi-edge-like line shape at 250 K, albeit with reduced \( \mu \)
weight, but this is not observed in the data.

The ARPES simulations shown in (a) and (b) of Fig.\ \ref{fig: blue bronze
  RS simulation} give a more detailed view.
Line shapes at and above \( \mathbf{k}_{F} \)
are most interesting. For both (a) and (b), the maximum \( \mu \) weight
occurs for \( \mathbf{k} \) somewhat greater than \( \mathbf{k}_{F} \).
For the moderate resolutions used in (a), the \( \mu \) weight at 101 \%
is significantly larger than half the peak height, but the data show
slightly less than half. The comparison becomes more problematic for high
resolutions used in (b). In this case, the peak occurs \emph{at} \( \mu \)
for 91 \%, and disappears quickly after that. Experimentally, however, the
\( \mu \) weight is never greater than half the peak height and the line
shape after crossing is nearly the same as that at the crossing. In
addition, notice that the large line width reduction from (a) to (b) is not
observed in the data.

\subsubsection*{LL}

In this section, we compare the blue bronze data with line shapes for a
spin-independent TL model \cite{Meden:1992:PRB}, in the same fashion
as was done in Section \ref{sec: li bronze}.  First, we examine whether
the experimental data show the power law predicted by the LL theory. This
can be done by examining the data of Fig.\ \ref{fig: blue bronze angle
  integ. collection} in a log-log plot, and identifying the region where
the plot is linear. As we noted before, this region is expected to start
at a finite binding energy, determined by \( t_{\perp } \), \( T \) and \(
\Delta E \)\@. We estimate an upper bound of \( t_{\perp } \) to be \(
\approx 30 \) meV \cite{tperp-estimate}.  The data of Fig.\ \ref{fig: blue
  bronze angle integ. collection} show power law behavior, \( \alpha
=0.5-0.8 \), starting from energy \( \approx \max (2k_{B}T,\Delta
E/2,t_{\perp }) \) to 150--200 meV\@. The variation of the \( \alpha \)
value seems to correlate with \( T \), but may also have correlations with
other factors such as angular resolution and sample. We choose the value
of \( \alpha =0.7 \) obtained from our data taken at 300 K -- the farthest
from the phase transition -- and having the largest angle acceptance.

\begin{figure}
  {\centering
    \resizebox*{0.95\columnwidth}{!}{\includegraphics{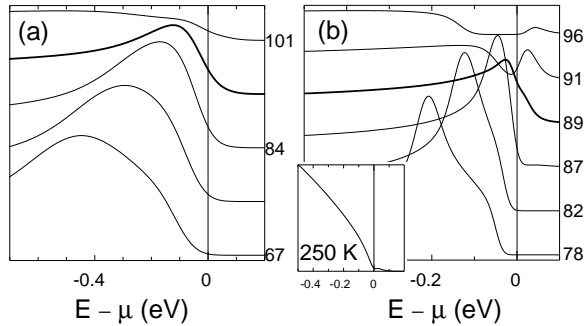}} \par}

\caption{Simulation of LL line shape, in a similar fashion as
  Fig.\ \ref{fig: blue bronze RS simulation}.
\label{fig: blue bronze TL simulation}}
\end{figure}

Next we compare the ARPES data with the calculated TL model line shapes of
Fig.\ \ref{fig: blue bronze TL simulation}. The parameters used for this
TL model are \( \alpha =0.7 \), \( \hbar v_{F}=0.98 \) eV\AA, and \(
r_{c}=0.1 \)\AA. The \( v_{F} \) value was chosen in order to reproduce
the dispersing peak with velocity 4.5 eV\AA.  The \( r_{c} \) value was
chosen so that the calculated spectral functions are well within the
validity limit of the universal LL behavior
\cite{Schonhammer:1993:PRB}.

The theoretical angle integrated spectrum in Fig.\ \ref{fig: blue bronze TL
  simulation} improves comparison with the data relative to that for the
RS theory, in that the TL theory predicts less \( \mu \) weight and no
Fermi edge. The amount of \( \mu \) weight has some uncertainty due to the
fact that the theory here does not include \( T \) and \( t_{\perp } \).
In its current form, the theory predicts less \( \mu \) weight in the
angle integrated spectrum than occurs in the data. Perhaps inclusion of \(
T \) and \( t_{\perp } \) would make the agreement better in this regard.

The comparison to the ARPES data is more involved. While the generally
lower \( \mu \) weight than in the CDW RS theory is in better agreement
with data, it is difficult to identify some key features of the
$\mathbf{k}$-resolved theory in the data. The spin edge singularities,
which provided an interpretation of the leading edges in the Li purple
bronze line shapes, are hard to identify in the blue bronze data. The
charge edge singularity after $\mu$ crossing is also hard to see. The high
resolution simulation of (b) reveals discrepancies: 
the theory shows a peak \emph{above} \( \mu \)
after crossing and a greatly reduced line width before crossing, none of
which is observed in the data.

\subsection{Discussion \label{sec: blue bronze line shape discussion}}

The comparisons of the preceding section show that neither of the two
theories explains the ARPES line shapes satisfactorily.  The essential
findings are that (1) the absence of a Fermi edge (up to 313 K; see Fig.\ 
\ref{fig: blue bronze angle integ. collection}) is very hard to reconcile
with the CDW theory, (2) the higher resolution ARPES simulation for both
theories predict too much weight at \( \mu \) and too strong a \(
\mathbf{k} \) dependence for \( \mathbf{k}\geq \mathbf{k}_{F} \), and (3)
the edge line shapes in the LL theory are not identified in the data. The
single EDC shown by Fedorov \emph{et al}.\ \cite{Fedorov:2000:JPhys-CondMatt}
enable us to infer point (2) also from their data.

The severe disagreement of the high resolution data with theory needs
careful thinking. Let us recall from Section \ref{sec: TiTe2} that if
intrinsic line shapes are sharp enough, then it is possible to observe
peaks moving above \( \mu \), as is indeed the case for our TL line shape
simulation of Fig.\ \ref{fig: blue bronze TL simulation}.  That this
behavior is not observed in the data then implies that the intrinsic line
shape is not sharp enough. We have already noted in fact that the data do
not show significant line width reduction upon resolution improvement.
This implies that the ARPES line width is not resolution limited and is
very large -- a few hundred meV's. The origin of such a large line width
is an open question. A mundane explanation invoking a non-ideal surface
condition -- a mixture of mosaics or a warped surface -- seems unlikely,
because we observe two \( \mu \) crossings at the X point (Section
\ref{sec: blue bronze data}) and a sharp Laue diffraction pattern.

Underlying the reasoning in the preceding paragraph is the assumption that
the intrinsic line shape is not gapped. However, this assumption is
dubious for the blue bronze.  As noted first by Voit
\cite{Voit:1993:JPhys-CondMatt}, the normal state transport data shows
spin-charge separation in that the spin susceptibility shows gapped
behavior (\( \Delta =20 \) meV) while the resistivity shows metallic
behavior. Therefore, he suggested that the Luther-Emery (LE) model
\cite{Luther:1974:LE} gives a good description of the normal state of the
blue bronze.  In this model, certain backward scattering between electrons
is included, in addition to the forward scattering already included in the
TL model, and a gap opens up in the spin channel.  Because a single
particle excitation involves simultaneous excitations of spin and charge,
this spin gap appears in the single particle line shapes
\cite{Orgad:2000:finite-T-TL-and-LE,Voit:1998:EurPhysJB}.  Such a gap could
be a reason why the \( \mu \) weight does not increase further upon
resolution improvement.

The contrasting behaviors of the spin susceptibility and the resistivity
was recognized earlier by Pouget
\cite{Pouget:1985:bb-shallow-band-theory}, who proposed a simple
explanation within the one electron band theory. An essential component of
this explanation is a flat band \( 56 \) meV above \( \mu \), which is
thermally occupied as \( T \) increases. This flat band also was used in
an explanation of the \( T \) dependent CDW wavevector. Indeed, the band
calculation by Whangbo and Schneemeyer showed such a band near the \(
\Gamma \) point. If this scenario is right, then this shallow band should
be detectable in ARPES at high $T$, e.g.\ in the normal state.  However,
this band is neither reproduced by new local density approximation (LDA)
band calculations \cite{Kim:2000:bb-lda,Canadell:2000:bb-lda} nor observed
by ARPES.  Therefore, the more exotic explanation for the \( T \)
dependent susceptibility by Voit, discussed in the previous paragraph,
gains more credibility.  The \( T \) dependent CDW wavevector would then
require an alternate explanation as well.  Recently, Fedorov \emph{et
  al.}\ \cite{Fedorov:2000:JPhys-CondMatt} proposed a model in which \( T \)
dependent electron hopping integrals are responsible for the \( T \)
dependent CDW wavevector. However, the data presented by these authors are
insufficient to support the model because the data were taken along a
single line in the 2D BZ\@. In the model of the paper the 2D character of
the FS is essential, implying imperfect nesting.  Then the CDW wave vector
should \emph{not} be the same as the nesting vector along a single line.
Very qualitatively, the 2D character of the FS found in our \( \mu \)
intensity map \cite{Gweon:1996:JPhys-CondMatt} appears to be less than
envisioned in the model or predicted by band theory.  In our opinion
additional experiments and a further consideration of various models of
the \( T \) dependent CDW wavevector are merited.

\begin{figure}
  {\centering
    \resizebox*{0.95\columnwidth}{!}{\includegraphics{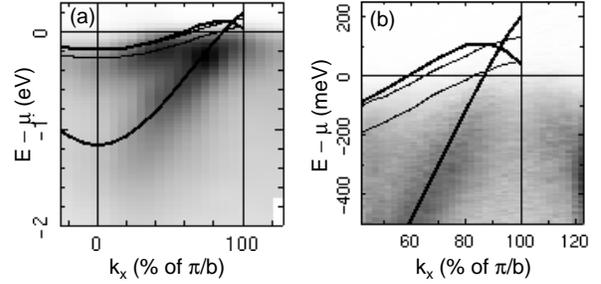}} \par}

\caption{Comparison of ARPES and band calculations for the blue
  bronze. Thin lines are tight-binding calculation
  \cite{Whangbo:1986:InorgChem} and thick lines are LDA calculation
  \cite{Kim:2000:bb-lda}. \label{fig: blue bronze ARPES and bands}}
\end{figure}

One of the characteristics of the repulsive TL model is that the charge
velocity is renormalized to be bigger than \( v_{F} \), in contrast to the
quasi-particle Fermi velocity smaller than \( v_{F} \) in the
case of the FL\@.
  It is therefore an interesting question how
the ARPES dispersions compare with the theory. Fig.\ \ref{fig: blue bronze
  ARPES and bands} shows the comparison.  We show two band calculation
results -- the tight binding theory by Whangbo and Schneemeyer
\cite{Whangbo:1986:InorgChem} and first-principles LDA theory by Kim
\emph{et al.}\ \cite{Kim:2000:bb-lda}. As noted previously
\cite{Gweon:1996:JPhys-CondMatt,Claessen:1995:JElecSpecRelPh}, 
the dispersion of band
A is a factor of 5 larger, compared to tight binding theory, and that of
band B is a factor of 2 larger. The new LDA theory, which should be more
accurate, is quite different from the tight binding theory.  The new LDA
band calculation is confirmed by that of another group
\cite{Canadell:2000:bb-lda}. Therefore the uncertainties in the magnitudes
of the one electron band dispersions seem finally to be gone.  The
dispersion of band A is in good agreement with that of the LDA theory and
that of band B is still about a factor of 2 larger.  This finding remains
as a piece of the whole blue bronze puzzle, and seems to require a better
understanding of the dependence of spin and charge velocities on $\alpha$
as we discussed in
Section \ref{sec: Li discussion}.

\section{Concluding Remarks \label{sec: conclusion}}

In this article, we have discussed three examples of ARPES line shape
studies of quasi-2D and quasi-1D samples showing FL and non-FL line
shapes.  The complex and intriguing line shapes of these prototypical
materials are not completely understood, and we strongly feel that they
are worth studying more both experimentally and theoretically, because
they connect to fundamental concepts of condensed matter physics.

Before concluding we comment on the common aspects of the LL parameters
for the bronzes, the large \( \alpha \) and energy scale. In the TL model
description we used an \( \alpha \) value of 0.7 (blue bronze) and 0.9 (Li
purple bronze).  Such an \( \alpha \) value may seem too large from the
point of view of the well-known 1D Hubbard model which has the maximum \(
\alpha \) value of 0.125. However, a better model to describe the Mo \( 4d
\) bands may be that of a free electron band with screened Coulomb
interactions. In this case, a coupled chain theory
\cite{Kopietz:1997:PRB}, evaluated for parameters appropriate for the
bronzes, shows that \( \alpha \approx 1 \) or larger is expected. In
addition, the Thomas-Fermi screening lengths for the bronzes are estimated
to be \( \approx \)0.7 \AA\ \cite{Thomas-Fermi}.  This means that the
universal form of the TL line shape used in Fig.'s \ref{fig:
  Li-PRL-line-shape-comparison} and \ref{fig: blue bronze
  TL simulation} are valid for
\( |\mathbf{k}-\mathbf{k}_{F}|\leq \) 1.1 \AA\( ^{-1} \) and \( |\omega
|\leq \) 0.7 eV, appropriately validating our model calculation.

One recent theoretical approach to the HTSC's is to consider them as
locally 1D quantum liquids. In essence, the basic model is the same as the
one considered here for the bronzes -- i.e.\ that of coupled 1D chains --
although the underlying physical Hamiltonians -- Hubbard-like or
free-electron-like -- are different. In fact, the phenomena that we
discussed in this article -- a pseudo-gap, a non-FL normal state, a
non-mean-field-like gap opening -- are also found in HTSC's. We
believe that our results on known quasi-1D systems can be used as a
standard in testing the 1D pictures for the HTSC's. In this context, it is
interesting that the non-mean-field-like $T$ dependence observed in the
blue bronze data (Fig.\ref{fig: blue bronze T dep. kF}) is reminiscent of
a recent theoretical result \cite{Carlson:2000:PRB} obtained for a
superconducting transition of coupled chains, in that both show a mere
intensity redistribution of ARPES spectra without a mean-field-like peak
shift as $T$ is lowered across the transition. However, for a further
elucidation of the blue bronze line shape, a similar theory designed for
the CDW is necessary. In such a theory, LL and CDW should be viewed as
tightly connected to, rather than independent of, each other. For example,
recently it was indicated how the $T$ dependence of the X-ray diffuse
scattering that arises from the CDW fluctuations in the quasi-1D organic
TTF-TCNQ family can be used to extract an LL exponent
\cite{Pouget:2000:JPhysIv}.

\section*{Acknowledgements}

G-HG and JWA thank J. Voit, K. Sch\"onhammer and S. Kivelson for very
useful discussions, and R. H. McKenzie for sharing his code for
calculating Sadovskii's spectral function.  The work at U. Mich.\ was
supported by the U.S. DoE under contract No.\ DE-FG02-90ER45416 and by the
U.S. NSF grant No.\ DMR-99-71611. Work at the Ames lab was supported by
the DoE under contract No.\ W-7405-ENG-82. The Synchrotron Radiation
Center was supported by the NSF under grant DMR-95-31009.
\vspace{8pt}

\noindent \emph{\normalsize \protect\( ^{1}\protect \)Current address: Advanced
  Light Source, Lawrence Berkeley National Lab., Berkeley CA 94720, USA}


\begin{thebibliography}{10}
\bibitem{SDW}
The same mechanism can also lead to the formation of spin density wave (SDW),
  especially if strong Hubbard-type electron-electron interactions exist. SDW's
  are not known to occur for the materials studied in this article, and so the
  SDW is not discussed.
\bibitem{edge-sharpening-limit}
The edge sharpening occurs down to temperature $T=\Delta E/(4k_B)$, below which
  the edge width is fixed by the energy resolution $\Delta E$\@.
\bibitem{different-vs}
Because the two nested bands have different slopes at $\mu$ in the blue bronze
  case, the new band dispersion is slightly more complicated. We leave this
  detail out in order to avoid unnecessary complication.
\bibitem{Thomas-Fermi}
This is basically the parameter $r_c$ used in the simulation of Fig.\ \ref{fig:
  Li-PRL-line-shape-comparison} and Fig.\ \ref{fig: blue bronze TL simulation}.
  For both simulations $r_c = 0.1$\AA\ was used. For $r_c \leq 0.7$\AA,
  however, the line shapes are practically independent of $r_c$.
\end{thebibliography}
\end{document}